\documentclass[fleqn,usenatbib]{mnras}
\usepackage{newtxtext,newtxmath}
\usepackage[T1]{fontenc}
\DeclareRobustCommand{\VAN}[3]{#2}
\let\VANthebibliography\thebibliography
\def\thebibliography{\DeclareRobustCommand{\VAN}[3]{##3}\VANthebibliography}
\usepackage{graphicx}
\usepackage{amsmath}
\usepackage{booktabs}
\usepackage{soul}

\title[47\,Tuc white dwarfs with \textit{JWST}]{\textit{James Webb Space Telescope} observations of the white dwarf cooling sequence of 47\,Tucan\ae\ }
\author[Salaris et al.]{
M.\,Salaris,$^{1,3}$
M.\,Scalco,$^{2}$
L.\,Bedin$^{2}$ and
S.\,Cassisi$^{3,4}$
\\
$^{1}$Astrophysics Research Institute, Liverpool John Moores University, 146 Brownlow Hill, Liverpool L3 5RF, UK\\
$^{2}$Istituto Nazionale di Astrofisica, Osservatorio Astronomico di Padova, Vicolo dell’Osservatorio 5, Padova I-35122, Italy\\
$^{3}$Istituto Nazionale di Astrofisica, Osservatorio Astronomico di Abruzzo, via M. Maggini, sn, 64100 Teramo, Italy\\
$^{4}$Istituto Nazionale di Fisica Nucleare, Sezione di Pisa, Universitá di Pisa, Largo Pontecorvo 3, 56127 Pisa, Italy\\
}

\date{Accepted 2025 June 24. Received 2025 May 24; in original form 2025 March 25}
 
\pubyear{\the\year{}}

\begin{document}
\label{firstpage}
\pagerange{\pageref{firstpage}--\pageref{lastpage}}
\maketitle

\begin{abstract}
We present a study of the white dwarf (WD) cooling sequence of the globular cluster 47\,Tucan\ae\  (47\,Tuc or NGC\,104) using deep infrared observations with the \textit{James Webb Space Telescope} (\textit{JWST}). By combining these data with ultra-deep optical imaging from the \textit{Hubble Space Telescope} (\textit{HST}) taken $\sim$12 years earlier, we derived precise proper motions (PMs) and isolated a clean sample of WD cluster members. We estimated the cluster’s age by comparing the observed WD cooling sequence luminosity function (LF), derived from \textit{JWST} photometry, with theoretical models, obtaining a value of $11.8\pm0.5$ Gyr, in agreement with 
main sequence turn-off ages, and ages determined from the  
masses and radii of two eclipsing binaries in the cluster. 
The age determined from the infrared LF is consistent with the optical LF based on the \textit{HST} photometry. However, small discrepancies exist between the shape of the observed and theoretical LFs. To investigate these differences, we tested the cooling times of WD models populating the bright 
part of the observed cooling sequence against a semi-empirical calibration from the literature, based on bright WDs in 47\,Tuc, finding agreement within less than 2$\sigma$. A more detailed analysis of dynamical effects and the impact of multiple stellar populations on the WD mass distribution in the observed field will be essential for addressing this discrepancy in future studies.
\end{abstract}

\begin{keywords}
globular clusters: general - globular clusters: individual: NGC\,104 - white dwarfs
\end{keywords}

\section{Introduction}\label{Section1}

The determination of the ages of Galactic globular clusters 
(GCs) has been a traditional goal of stellar astrophysics, because of its importance in setting a solid 
lower limit for the age of the universe, and in providing 
crucial insights into the process of formation of 
the Galaxy.

Milky Way-like galaxy haloes are predicted to 
be built up by the accretion of smaller galaxies plus 
a contribution from stars and clusters formed in situ  
\citep[see, e.g.,][]{abadi, font, zolotov} and although the 
event might destroy completely the accreted dwarf galaxy, some of its high-density GCs can survive and be added to the in situ halo GC population.

The identification of the accreted GCs provides therefore additional clues about the past evolution of the Milky Way, and it is indeed a hot topic in current GC research \citep[see, e.g.,][]{fb,mkh,tg, mass23, bk, aa}. 
In addition to 
information about the kinematics (when available) and chemical composition, age also plays an important role in determining whether a GC has originated in situ or has been accreted, because of 
distinctive age-metallicity relationships, 
as discussed by, e.g., \citet{fb}, \citet{leaman}, \citet{mkh}, \citet{callingham}. 

For most of the GCs investigated so far, age is determined 
essentially from their main sequence turn-off (TO) luminosity 
\citep[see, e.g.,][]{vdbages}, 
whilst for just a few clusters it has been possible to compare and confirm the TO age with the age independently 
determined from their white dwarf (WD) cooling sequence observed with the \textit{Hubble Space Telescope} (\textit{HST}) and more recently the   
\textit{James Webb Space Telescope} (\textit{JWST}), namely 
NGC~6397, M~4, and NGC~6752 
\citep[see, e.g.][]{2002ApJ...574L.155H, 2007ApJ...671..380H, 2009ApJ...697..965B, 2023MNRAS.518.3722B, 2024AN....34540039B, 2025AN....34640125B}.

The well-studied metal-rich GC 47\,Tucan\ae\  (47\,Tuc, NGC\,104) is the fourth GC whose age has been determined from both the TO and its CS, and in addition also from the mass-radius diagram of two 
eclipsing binaries populating the TO region of the cluster 
colour-magnitude diagram \citep[CMD -- see, e.g.,][]{vdbages, brogaard,eclbin}.
Recent age determinations from the TO and eclipsing binaries are consistent, around 12~Gyr, and locate 47\,Tuc firmly within the 
age-metallicity relation of in situ GCs.

The cluster age from its cooling sequence has delivered 
so far contradicting results. Using optical \textit{HST} data 
\citet{47tuchansen} have obtained 9.9$\pm$0.7~Gyr (95\% confidence level), 
which makes 47\,Tuc too young for the age-metallicity relation of in situ clusters. On the other hand, 
\citet{47tucgb} determined a cooling sequence age around 12 Gyr using a different 
set of WD 
models, consistent with TO and eclipsing binary ages.
In between these two results, \citet{campos} found $10.95^{+0.21}_{-0.15}$ and $11.31^{+0.36}_{-0.17}$ when using 
the distance modulus and reddening derived from fitting with WD 
models the bright, age-independent  
part of the CS, and the distance and reddening obtained by \citet{dotterdist} from fitting models to the cluster main sequence, respectively.

For this reason,  we have re-investigated the cooling sequence age of 47\,Tuc by taking advantage of recent state-of-the-art WD evolutionary models 
\citep[][]{bastiiacwd} that include 
among others, convective coupling between envelope and electron degenerate layers, CO phase separation upon crystallization and diffusion of $^{20}$Ne in the liquid phase (and also tested the effect of $^{20}$Ne distillation upon crystallization), an improved distance 
from two cluster's eclipsing binaries \citep{eclbin}, 
and new \textit{JWST} observations of the cluster CS.

The plan of the paper is as follows. Section~\ref{Section2} 
describes the \textit{JWST} data, the reduction process and the proper motion selection, whilst Sect.~\ref{Section3} discusses the artificial star tests.
Section~\ref{Section4} presents the WD luminosity function 
used in Sect.~\ref{age} to determine the cluster age. A summary and conclusions follow in Sect.~\ref{conclusions}.

% link
% https://ui.adsabs.harvard.edu/abs/2013ApJ...778..104R/abstract

\section{Data set and reduction}\label{Section2}
The data used in this study were presented in \citet{2025A&A...694A..68S}, to which we refer the reader for a detailed description of the dataset and reduction process.

Briefly, the dataset originates from \textit{JWST} program GO-2559 \citep{2021jwst.prop.2559C}, and observations were conducted using the Near Infrared Camera \citep[NIRCam,][]{2023PASP..135b8001R} over two epochs ($\sim$2022.7 and $\sim$2023.4) with the F150W2 filter for the Short Wavelength (SW) channel and the F322W2 filter for the Long Wavelength (LW) channel. As in \citet{2025A&A...694A..68S}, we rely only on the first epoch's photometry for this study, as it provides deeper and higher-quality data.

Data reduction of the images involved a combination of first- and second-pass photometry \citep[see also][for a description of the \textit{JWST} data reduction process]{2024AN....34540039B, 2024A&A...689A..59S,2022MNRAS.517..484N, 2023MNRAS.525.2585N, 2023MNRAS.521L..39N, 2023AN....34430006G, 2023ApJ...950..101L, 2024PASP..136c4502L}, using the software \texttt{KS2} for the second-pass \citep[see][and references therein for a description of \texttt{KS2}]{2017ApJ...842....6B,2018ApJ...853...86B,2018MNRAS.481.3382N,2018ApJ...854...45L,2022ApJ...934..150L,2021MNRAS.505.3549S}. 

The astrometry was anchored to the absolute reference frame provided by \textit{Gaia} Data Release 3 \citep[DR3;][]{2016A&A...595A...1G,2023A&A...674A...1G}, and the photometry was calibrated to the Vega-magnitude photometric system following the prescription of \citet{2005MNRAS.357.1038B}.

We selected a sample of well-measured stars by applying quality-based selection criteria using parameters provided by \texttt{KS2}, such as the QFIT, which quantifies the accuracy of the PSF fitting, and RADXS parameters, which measure how closely a source’s shape resembles the PSF. The RADXS parameter is particularly useful for distinguishing between point-like (stellar) and extended (non-stellar, e.g., galaxies or blends) sources \citep[see][]{2008ApJ...678.1279B,2009ApJ...697..965B}. We refer to \citet{2021MNRAS.505.3549S} for a comprehensive description of these parameters.  In this study, we retained only sources with absolute RADXS values smaller than 0.075 in the F150W2 filter and smaller than 0.15 in the F322W2 filter. For the QFIT parameter, we required positive values in both filters, as negative values typically indicate unreliable PSF fitting. In this study, the adopted thresholds are broader than those used in \citet{2025A&A...694A..68S}, which were optimized for main-sequence stars. Because WDs are intrinsically faint, and faint stars tend to have larger photometric uncertainties and degraded shape measurements, stricter cuts would unnecessarily exclude a significant number of valid WD candidates. We therefore adjusted the selection criteria to strike a balance between data quality and completeness for the WD population, which is the main focus of this work.

In \citet{2025A&A...694A..68S}, proper motions (PMs) were measured using the displacements between the two \textit{JWST} GO-2559 epochs. While the use of \textit{JWST} data provides exceptional depth for PM measurements 
% R: added
of red objects (invisible in the optical), 
the short temporal baseline between the two \textit{JWST} epochs ($\sim$7\,months) limits the accuracy of PM estimates for fainter stars, such as the WDs targeted in this study. This baseline was adequate for the analysis in \citet{2025A&A...694A..68S}, but for this work, a longer temporal baseline is essential for an accurate measure of the PMs of faint 
% R: added
--bluer-- 
stars and to disentangle them from objects in the Small Magellanic Cloud (SMC).

\begin{centering} 
\begin{figure}
 \includegraphics[width=\columnwidth]{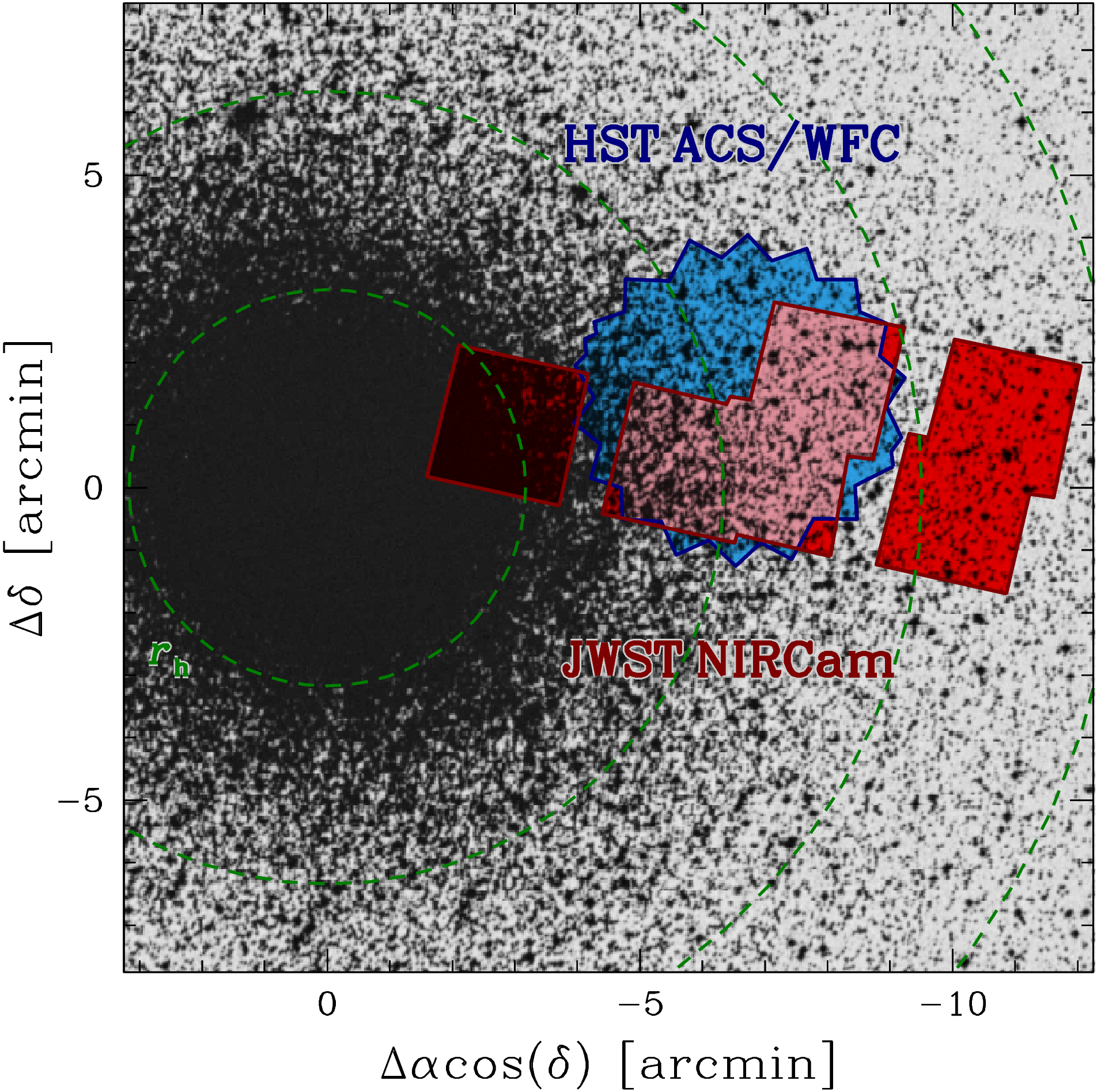}
 \caption{The NIRCam \textit{JWST} field from the first epoch of the GO-2559 programme (red) and the ACS/WFC \textit{HST} field from the GO-11677 programme (blue) overlaid on a DSS image of 47\,Tuc. The overlap region between the two datasets is highlighted in magenta. Units are in arcminutes from the cluster's centre. The green dashed circles represent the half-light radius ($r_{\rm h}=3^\prime_{\cdot}17$; \citealt{1996AJ....112.1487H,2010arXiv1012.3224H}), with additional circles marking $2\,r_{\rm h}$, $3\,r_{\rm h}$, and $4\,r_{\rm h}$.} 
 \label{FOV} 
\end{figure} 
\end{centering}

To address this limitation, we include \textit{HST} images from program GO-11677 \citep{2009hst..prop11677R}, taken between 15 January and 1 October, 2010 (epoch $\sim$2010.4). These \textit{HST} images partially overlap with the \textit{JWST} GO-2559 FOV and provide a significantly longer temporal baseline, enabling more precise PM measurements. They were collected by the Advanced Camera for Surveys (ACS) with the F606W and F814W filters. The reduction of the \textit{HST} images was performed following the same first- and second-pass photometry procedures applied to the \textit{JWST} data \citep[see][for additional details on \textit{HST} data reduction]{2021MNRAS.505.3549S}. Figure\,\ref{FOV} shows the positions of our \textit{JWST} and \textit{HST} datasets within the field of view (FOV), overlaid on a Digital Sky Survey\footnote{\href{https://archive.eso.org/dss/dss}{https://archive.eso.org/dss/dss}} (DSS) image of 47\,Tuc.

\begin{centering} 
\begin{figure*}
 \includegraphics[width=\textwidth]{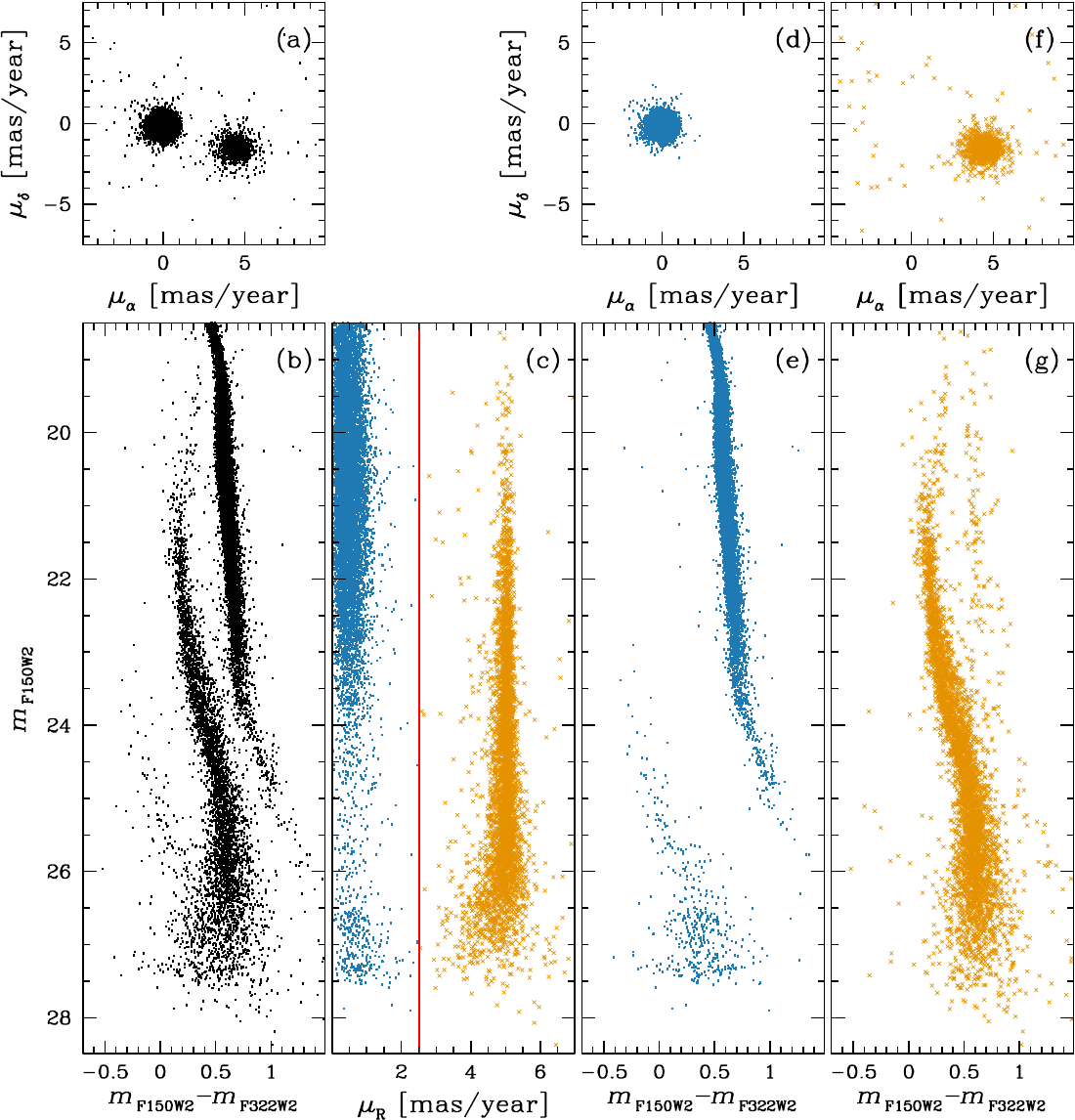}
 \caption{Custer membership selection based on PMs. The figure displays only well-measured stars that meet the \texttt{KS2} quality selection criteria and have detectable PMs. (a) VPD. (b) $m_{\rm F150W2}$ versus $m_{\rm F150W2}-m_{\rm F322W2}$ CMD, focused on the low-MS, the SMC MS and the WD CS. (c) $m_{\rm F150W2}$ magnitude versus the one-dimensional PM ($\mu_{\rm R}$). The red line separates cluster members from field stars. (d)-(e) VPD and $m_{\rm F150W2}$ versus $m_{\rm F150W2}-m_{\rm F322W2}$ CMD for stars that passed the PM selection. (f)-(g) the same diagrams for stars that did not pass the PM selection. In panels (c) through (g), sources that passed the PM selection are represented by blue dots, while the others are depicted as orange crosses.} 
 \label{pm} 
\end{figure*} 
\end{centering}

Proper motions were calculated as the displacements of stars common to both the \textit{JWST} and \textit{HST} datasets, divided by the temporal baseline ($\sim$12 years). The resulting PMs for our selected sample of well-measured stars with detectable PMs are shown in Fig.\,\ref{pm}. Panel\,(a) shows the vector-point diagram (VPD), while panel\,(b) shows the $m_{\rm F150W2}$ versus $m_{\rm F150W2}-m_{\rm F322W2}$ CMD, focused on the low-MS, the SMC MS and the cluster CS. In the VPD, two distinct groups of stars are evident: the cluster members, centred at (0,0), and the SMC stars, located around (5,$-$0.5). The distribution of points in this study is significantly narrower than that in \citet[][see Fig.\,3]{2025A&A...694A..68S}, owing to the larger temporal baseline used here.

Panel\,(c) displays the one-dimensional PM ($\mu_{\rm R}$), obtained by combining the PM components in quadrature, plotted against $m_{\rm F150W2}$. The separation between cluster members and SMC stars is particularly evident along the MS and remains clear for stars on the WD CS. Most 47\,Tuc cluster members are concentrated below $\mu_{\rm R} < 2.5$ mas yr$^{-1}$, whereas SMC stars cluster around $\mu_{\rm R} \sim 5$. A PM-based selection criterion, indicated by a red line, was defined to separate cluster members from SMC objects.

Panel\,(d) displays the VPD for stars that passed the PM selection, while panel\,(e) presents the corresponding CMD, where the WD cooling sequence of the cluster is visible clearly. Finally, panels\,(f) and (g) show the VPD and CMD, respectively, for stars that did not satisfy the PM selection criteria. From panel\,(f), it is evident that a small number of stars remain near the cluster’s PM locus after applying the selection criterion, suggesting a residual level of contamination in the cluster sample. We estimated the contamination rate using artificial stars (see Section\,\ref{Section3}) by calculating the fraction of artificial stars rejected solely based on their PMs, relative to the total number of artificial stars. This analysis indicates an estimated contamination rate of $\sim$8\%.

Figure\,\ref{cmd} in the appendix presents a set of CMDs of the WD cooling sequence of 47\,Tuc, using various combinations of the four \textit{HST} and \textit{JWST} filters employed in this study. The CMDs include all well-measured sources that passed the PM selection described in Fig.\,\ref{pm}.

\section{Artificial stars}\label{Section3}
We performed artificial star tests to evaluate the completeness and estimate the photometric errors in our sample. A total of 10$^5$ artificial stars were generated, uniformly distributed across the overlapping FOV between the \textit{JWST} and \textit{HST} datasets. The F150W2 magnitudes of these  artificial stars were uniformly sampled within the range $22 < m_{\rm F150W2} < 30.5$. The corresponding magnitudes in the F322W2, F606W, and F814W filters were assigned based on fiducial lines manually defined on the $m_{\rm F150W2}$ versus $m_{\rm F150W2}-m_{\rm F322W2}$, $m_{\rm F150W2}$ versus $m_{\rm F606W2}-m_{\rm F150W2}$ and $m_{\rm F150W2}$ versus $m_{\rm F814W2}-m_{\rm F150W2}$ CMD, respectively. These fiducial lines trace the WD CS, extending to the apparent faint end of the sources, and extrapolated to even fainter magnitudes. 

The artificial stars were generated, detected, and measured using \texttt{KS2}, following the same procedures applied to the real stars. Artificial PMs were generated by separately processing the artificial stars for the \textit{JWST} and \textit{HST} datasets and evaluating their displacements between the two epochs.\footnote{
%%%
Spurious positional offsets due to noise, especially in faint stars, can affect PM measurements and membership selection. Since artificial stars are generated with identical positions in both epochs, any observed displacement is solely ascribed to noise. This approach allows us to quantify this effect, and include it in the completeness evaluation.
%%%
}

We followed the methodology described in \citep[][Sect.\,2.3]{2009ApJ...697..965B} to correct for systematic errors between input and output magnitudes in both real and artificial sources. We found these corrections negligible (<0.1 mag) for the \textit{JWST} data down to the faintest magnitudes studied, therefore we applied them only to the \textit{HST} photometry.

An artificial star is considered successfully recovered if the difference between its input and output positions is less than 1 pixel, the difference in magnitudes is within 0.75 (equivalent to $\sim$\,2.5log\,2) in both filters and if it passes the same selection criteria applied to real stars. The completeness is then calculated as the ratio between the recovered and the injected artificial stars as a function of the magnitude.

\begin{centering} 
\begin{figure}
 \includegraphics[width=\columnwidth]{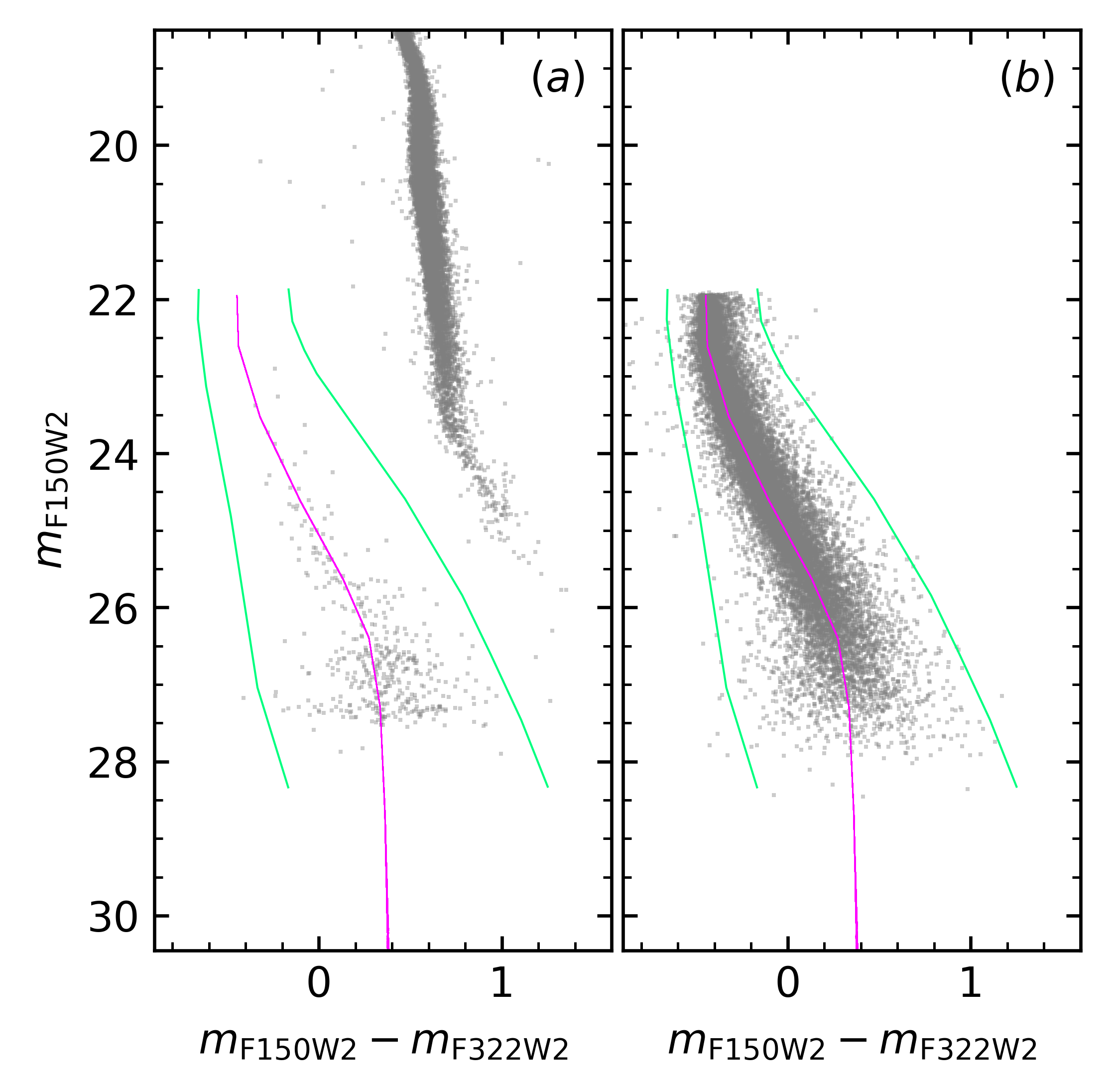}
 \caption{(a) $m_{\rm F150W2}$ versus $m_{\rm F150W2}-m_{\rm F322W2}$ CMD for the real stars.
 % (grey points). 
 (b) Same as (a) but for the recovered ASs.
 %(grey points). 
 In both panels, magenta points represent the injected ASs. The two green fiducial lines are used to isolate the stars along the WD CS.} 
 \label{fid} 
\end{figure} 
\end{centering}

Figure\,\ref{fid} shows the $m_{\rm F150W2}$ versus $m_{\rm F150W2}-m_{\rm F322W2}$ CMD for the real stars (points in panel a) and the recovered artificial stars (points in panel b), with the injected artificial stars represented in magenta in both panels. To isolate the WDs, we manually defined two fiducial lines (shown in green) that enclose the WD cooling sequence in both panels. These fiducial lines will be used in the following to define the sample of WDs for the 
luminosity function (LF) study.

We employed the artificial stars to estimate photometric errors as a function of magnitude for each of the four filters. To do this, the recovered artificial stars were divided into 0.5-magnitude bins for each filter. Within each bin, we computed the 2.5\,$\sigma$-clipped median of the distribution of differences between the injected and recovered magnitudes. The final $\sigma$ value from the clipping process was adopted as our estimate of the photometric error.

\section{Luminosity function}\label{Section4}
\begin{centering} 
\begin{figure*}
 \includegraphics[width=0.325\textwidth]{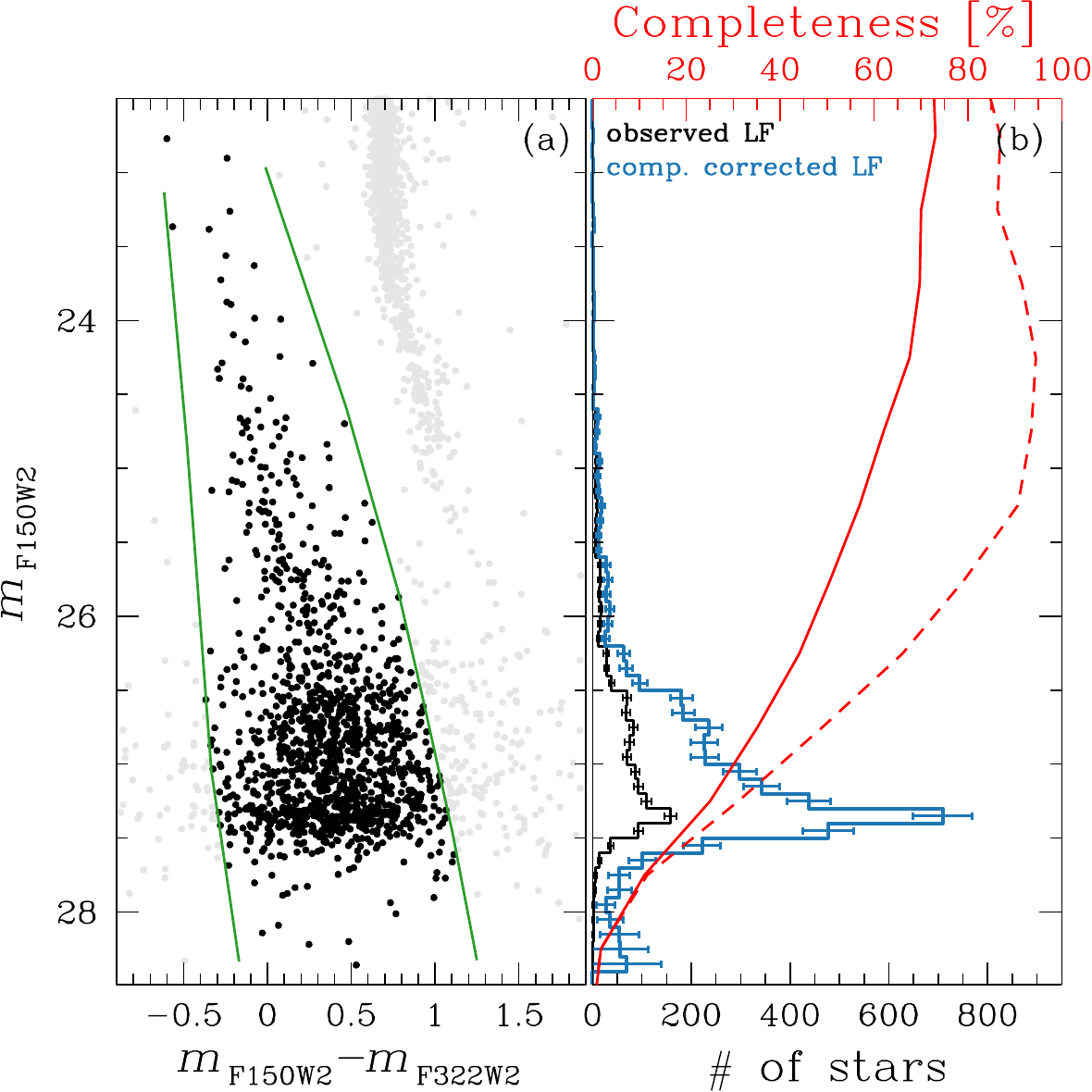}
 \includegraphics[width=0.325\textwidth]{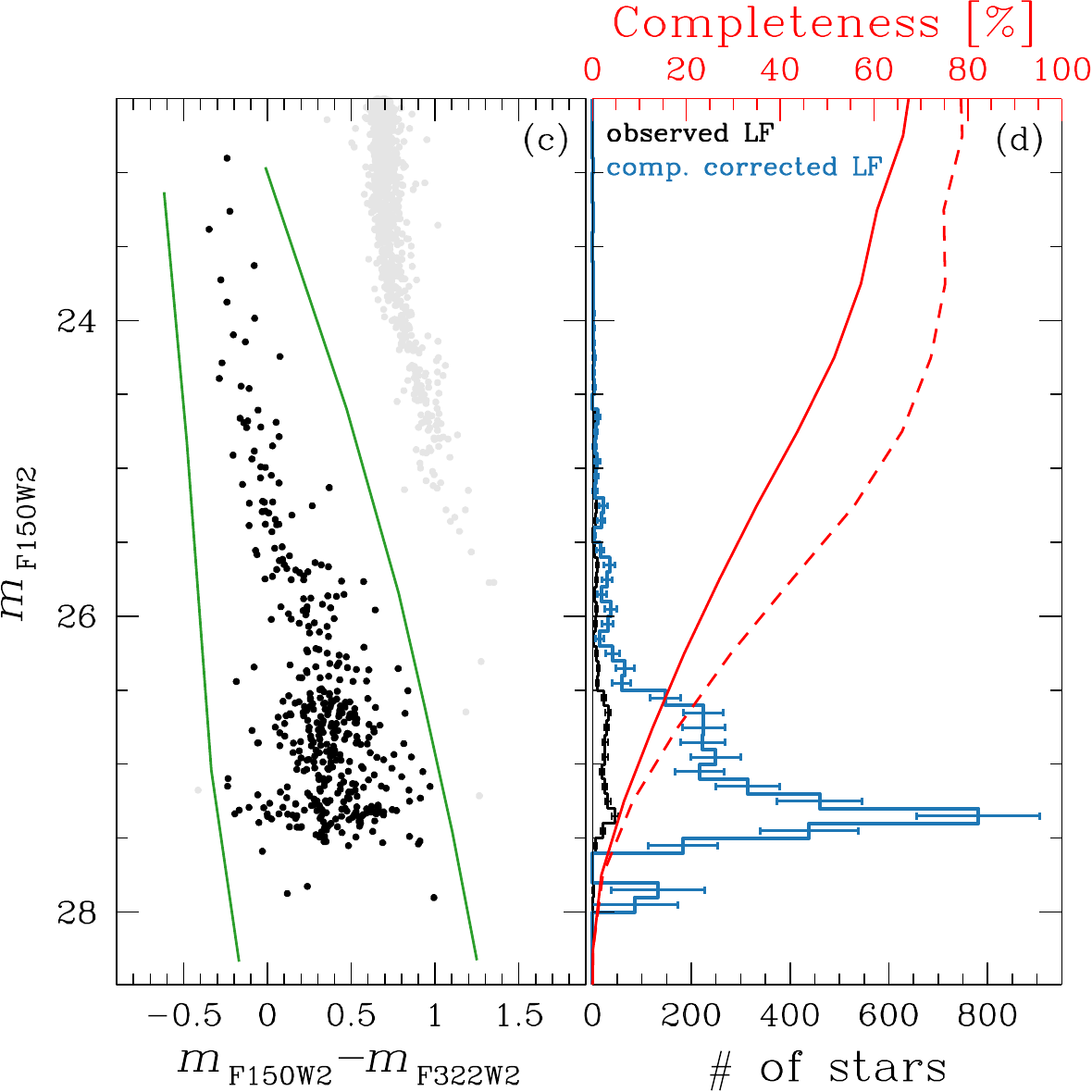}
 \includegraphics[width=0.33\textwidth]{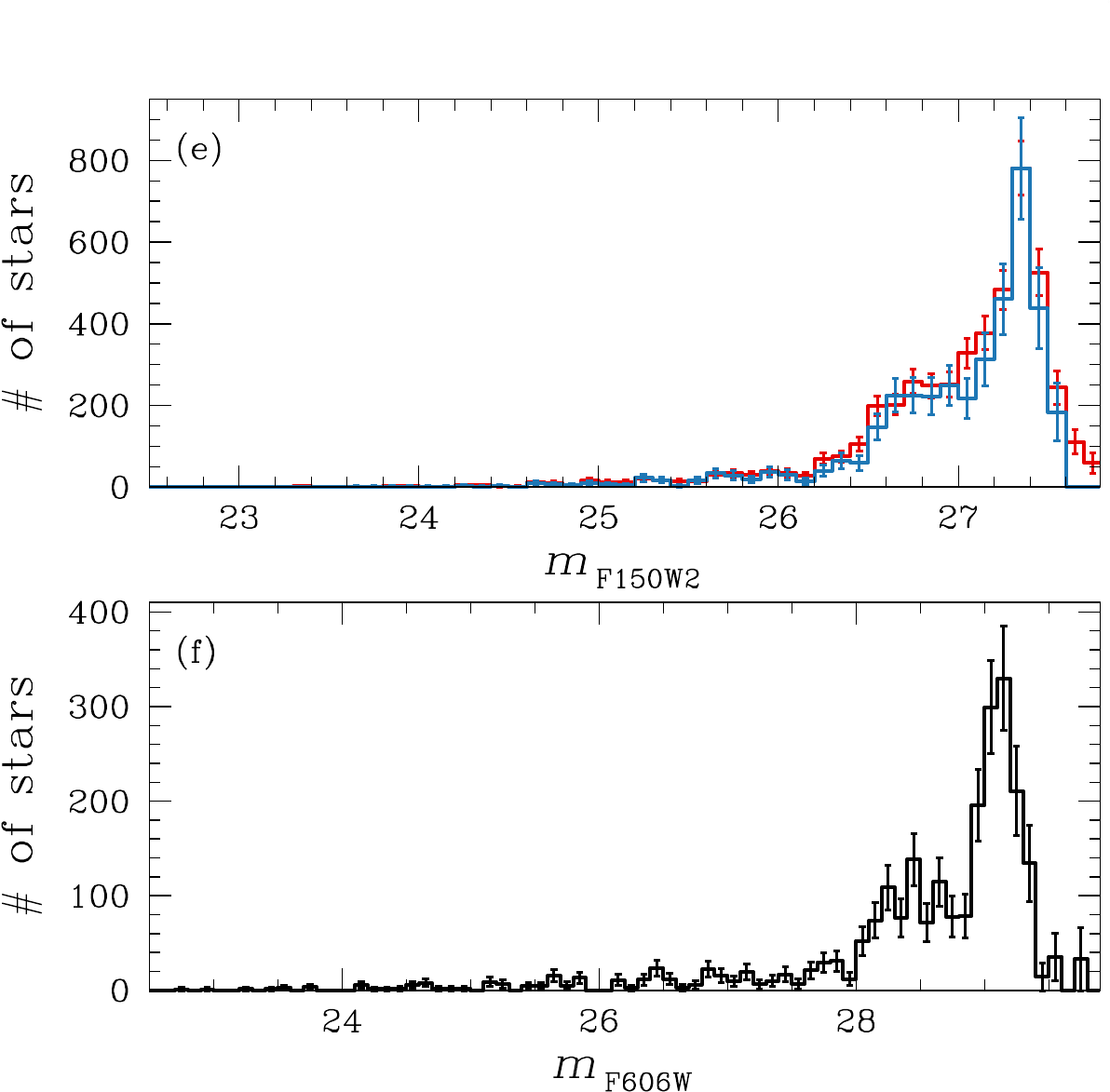}
 \caption{Luminosity function of the cooling sequence of 47\,Tuc. (a) $m_{\rm F150W2}$ versus $m_{\rm F150W2} - m_{\rm F322W2}$ CMD of stars that pass only the PMs selection. The green line defines the region used to select WDs for the evaluation of the LF. The selected WDs are shown in black, while all other sources are shown in grey. (b) Observed (black) and completeness-corrected (blue) LF of the WDs selected in (a), with the corresponding errors shown as error bars. The standard completeness and $c_g$ as a function of the $m_{\rm F150W2}$ magnitude are represented as solid and dashed red lines, respectively, with the corresponding axis shown at the top of the panel. (c)-(d) Same as panels (a) and (b), but for stars passing both the PMs and the selections based on \texttt{KS2} photometric quality parameters. (e) Comparison of the LFs derived from the two samples: the LF for stars that pass only the PMs selection is shown in red, while the LF for stars that pass both the PMs and photometric selections is shown in blue. Note that the red LF and its errors have been normalized to align the peaks of both LFs. (f) LF of 47\,Tuc derived using optical \textit{HST} photometry.} 
 \label{lf} 
\end{figure*} 
\end{centering}

In this section, we present the LF of the WD cooling sequence in 47\,Tuc, based on \textit{JWST} data. The analysis is shown in Fig.\,\ref{lf}. To evaluate the LF, we initially excluded the selection based on the \texttt{KS2} photometric quality parameters and considered a sample of stars selected only using the PM criteria shown in Fig.\,\ref{pm}.

Panel\,(a) shows the $m_{\rm F150W2}$ versus $m_{\rm F150W2} - m_{\rm F322W2}$ CMD for these sources, where the green lines, previously introduced in Fig.\,\ref{fid}, define the WD selection region. Objects within this region are represented in black, while all other sources are shown in grey.

Panel\,(b) presents the observed LF of the selected WDs as a black histogram. The completeness corresponding to the selected sample, as a function of $m_{\rm F150W2}$ magnitude, is shown as a red solid line, with its corresponding values shown on the secondary axis at the top of the panel. The red dashed line represents $c_g$, the completeness values restricted to the darker usable areas for detecting faint sources \citep[see][for further details]{2008ApJ...678.1279B,2009ApJ...697..965B}. The completeness-corrected LF is displayed as a blue histogram. Error bars indicate uncertainties: for the observed LF, they correspond to Poisson errors, while for the completeness-corrected LF, they result from uncertainty propagation. 

We then repeated the analysis by adding a selection based on the \texttt{KS2} photometric quality parameters. The results are displayed in panels\,(c) and (d), together with the corresponding completeness.

The completeness-corrected LF in panel\,(b) is noisier compared to the one in panel\,(d) due to higher contamination from poorly measured or non-member sources. On the other hand, the LF presented in panel\,(d) is cleaner and more reliable but suffers from significantly lower completeness near the peak ($m_{\rm F150W2} \sim 27.5$), reaching values as low as 5\%, below the thresholds typically considered reliable.

Panel\,(e) shows the comparison between the two completeness-corrected LFs: the one obtained in panel\,(b) is shown in red, while the one from panel\,(d) is shown in blue. The red LF and its errors have been normalized to align the peaks of both LFs. As shown, the two LFs are completely consistent with each other. While the completeness for the blue LF is very low below the LF peak ($m_{\rm F150W2} > 27.5$), with values approaching 0 (see panel\,d), the completeness for the red LF in this region remains acceptable, around 20\% (see panel\,b). Nonetheless, the drop below the peak is also observed in the red LF, confirming that the drop is real and not an artefact of the stricter selection criteria. The consistency between the two LFs, supports the validity of the blue LF, even at lower completeness levels.

For the age determination discussed in the next section, we will focus exclusively on the blue LF shown in panel\,(e), as it is the most reliable due to its higher precision and reduced contamination.

For comparison, and to use as a consistency check of the ages derived from the \textit{JWST} LF, we also derived the WD LF from the \textit{HST} photometry in our dataset. We selected the same sample of stars used to estimate the blue LF in panel\,(e), applying both the \texttt{KS2} photometric quality parameters and the PM-based membership selection. However, in this case, we defined the WD cooling sequence in the $m_{\rm F606W}$ versus $m_{\rm F606W} - m_{\rm F814W}$ CMD using two fiducial lines, following the approach illustrated in Fig.\,\ref{fid}. The completeness-corrected LF is presented in panel\,(f) of Fig.\,\ref{lf}. As seen in the figure, the overall shape of this LF closely resembles that of the \textit{JWST}-derived LF.

The infrared LF used for the WD age determination and the optical 
counterpart from \textit{HST} data are publicly available on our website\footnote{\href{https://web.oapd.inaf.it/bedin/files/PAPERs\_eMATERIALs/JWST/Paper\_47Tuc\_WDCS/}{https://web.oapd.inaf.it/bedin/files/PAPERs\_eMATERIALs/JWST/ \\ Paper\_47Tuc\_WDCS/}} and are tabulated in Tables\,\ref{tab:jwst} and \ref{tab:hst} in the appendix.

\section{The age of 47\,Tuc from its WD cooling sequence}\label{age}

To determine the age of 47\,Tuc from its cooling sequence we have taken advantage of  
the BaSTI-IAC hydrogen-atmosphere WD models\footnote{See \citet{davis} for the apparent dearth of He-atmosphere WDs in globular clusters. In the Appendix we show the effect 
of including helium-atmosphere objects in our analysis.} and isochrones \citep[see][for a detailed description of the models]{bastiiacwd} calculated with the \citet{opadeg} electron degeneracy opacities, for a progenitor metallicity corresponding to [Fe/H]=$-$0.7\footnote{The WD models can be found at \url{http://basti-iac.oa-abruzzo.inaf.it/}.}, consistent with [Fe/H]=$-$0.66$\pm$0.04 derived by \citet{gratton}, [Fe/H]=$-$0.67$\pm$0.05 obtained by \citet{carretta04}, [Fe/H]=$-$0.768$\pm$0.016$\pm$0.031 
(random and systematic errors) measured by 
by \citet{carr09}, 
and $-$0.70$\pm$0.01$\pm$0.04 (random and systematic errors) determined 
by \citet{koch} from high-resolution spectroscopy. In the calculation 
of the isochrones we have employed progenitor lifetimes 
from the BaSTI-IAC $\alpha$-enhanced ([$\alpha$/Fe]=0.4) 
models by \citet{bastiiacaen}, and the WD initial-final mass 
relation by \citet{cummings}. The WD models include, among others,  
the effect of 
convective coupling between envelope and electron degenerate layers, 
latent heat release and CO phase separation upon crystallization, plus $^{22}$Ne diffusion in the liquid phase \citep[see][for details]{bastiiacwd}. 
At this metallicity the diffusion of $^{22}$Ne increases the cooling time 
by at most 200-300~Myr for the faintest objects.

We have also calculated with the same code of 
the WD computations some test models that in addition include $^{22}$Ne distillation during crystallization  
\citep[see][]{blouindistillation, bedard24, bastidistillation}, and found that 
at the cluster's relatively low metallicity 
this process has a negligible impact on the energy budget and cooling times.

\begin{centering} 
\begin{figure}
 \includegraphics[width=\columnwidth]{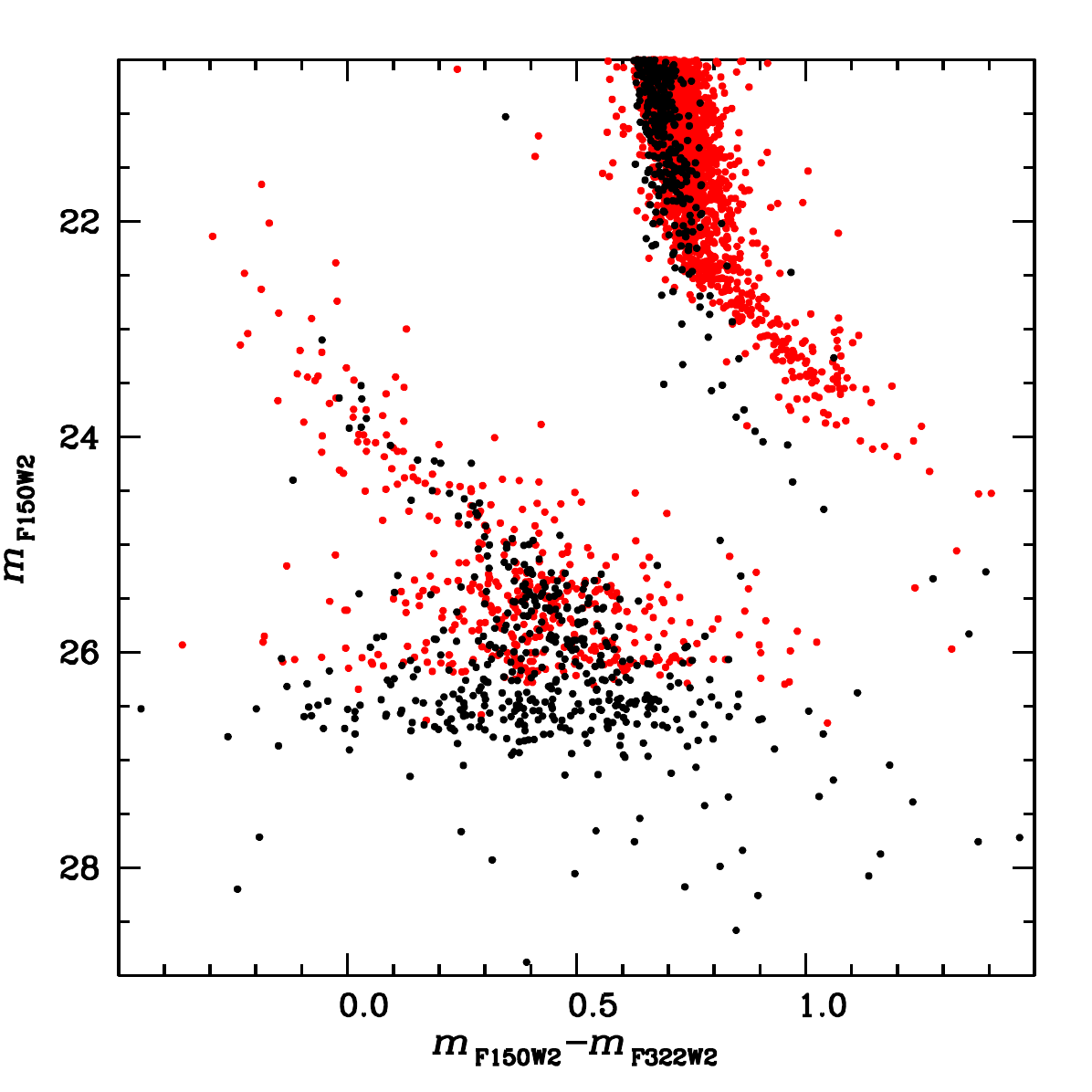}
 \caption{CMD of NGC~6397 cooling sequence \citep[black dots -- from][]{2024AN....34540039B} compared to the 47\,Tuc cooling sequence   
 (red dots -- the same sequence of Fig.~\ref{cmdiso}) shifted in magnitude and colour to the same distance and reddening of 
 NGC~6397.} 
 \label{cmdshift} 
\end{figure} 
\end{centering}

Figure~\ref{cmdshift} 
displays the CMD of the cooling sequence and the very low main sequence stars in NGC~6397 \citep[from][]{2024AN....34540039B} and 47\,Tuc, this latter CMD shifted in magnitude and colour to the same distance and reddening of NGC~6397.
Here and in the rest of this work we have employed as reference values for the distance modulus of 47\,Tuc 
($m-M)_0$=13.29, taken from the analysis of two cluster's eclipsing binaries by \citet{eclbin}\footnote{Parallax and kinematic distance determinations from $Gaia$ Data Release 3 are consistent with the eclipsing binary distance \citep{distances}.}, and for the 
reddening $E(B-V)$=0.024
\citep{gratton}, using the extinction ratios $A_{\lambda}/A_V$ we employed in \citet{2024AN....34540039B}, whilst 
for NGC~6397 we have used ($m-M)_0$=11.96 and $E(B-V)$=0.18 
\citep[see][]{2024AN....34540039B}. 

As found in optical \textit{HST} filters 
by \citet{richer} and \citet{47tuchansen}, also in the infrared the overall morphology of 
47\,Tuc cooling sequence is consistent with that of the much more metal-poor 
GC NGC~6397 ([Fe/H]$\sim -$2.0). 
Whilst the lower main sequence of 47\,Tuc is redder because of its higher metallicity, the two cooling sequences overlap nicely. The termination of the shifted 47\,Tuc cooling sequence is however brighter than the NGC~6397 counterpart, like in the optical, suggesting an age younger than the $\sim$13~Gyr found by \citet{2024AN....34540039B} for NGC~6397 using the same models (for both WDs and their progenitors) employed in this analysis.

\begin{centering} 
\begin{figure}
 \includegraphics[width=\columnwidth]{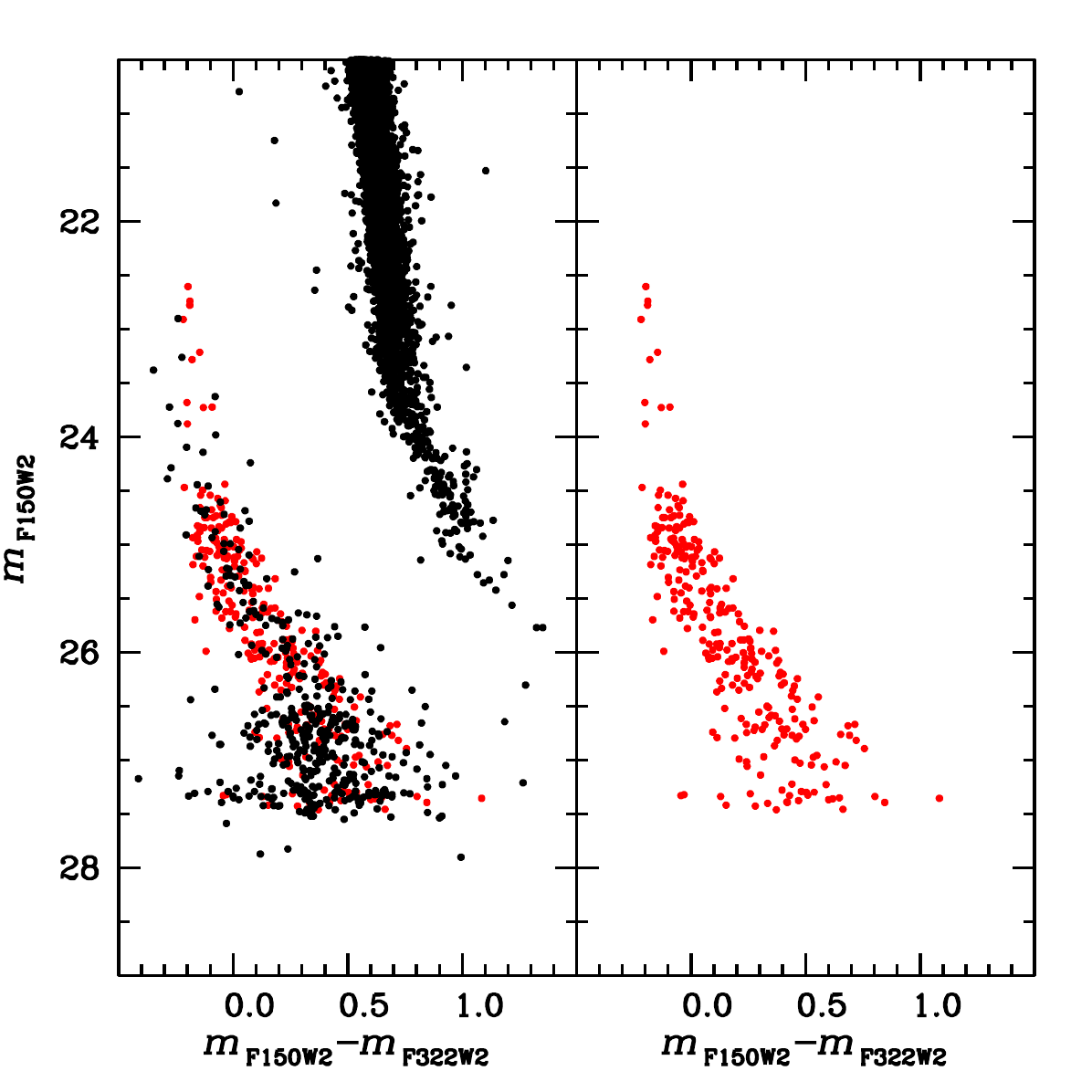}
 \caption{The right panel displays a synthetic CMD of a 12~Gyr old WD population with [Fe/H]=$-$0.7 progenitors, photometric errors and completeness fractions as those derived from the artificial star tests, shifted to 47\,Tuc  distance modulus and reddening (see text for details). The left panel shows the same synthetic WDs (red dots) together with the observed CMD of 47\,Tuc cooling sequence and very-low main sequence (black dots).} 
 \label{cmdiso} 
\end{figure} 
\end{centering}

Figure~\ref{cmdiso} shows a first qualitative comparison between 
the $m_{\rm F150W2}$ versus $m_{\rm F150W2} - m_{\rm F322W2}$ synthetic CMD generated from a 12~Gyr WD isochrone, and the observed CS. We use the cooling sequence in panel (c) of Fig.\,\ref{lf}, consistent with the LF of panel (d) employed in the rest of this analysis, as discussed in the previous section. 

The synthetic cooling sequence has been calculated by 
first drawing progenitor masses along the isochrone using a power-law mass function (MF) $dN/dM$ $\propto$ $M^{\alpha}$, with $\alpha=-$2.35; the corresponding WD and its F150W2 and F322W2 magnitudes were then determined by quadratic interpolation along the isochrone. 
We then added our adopted reference distance modulus and extinction values to
these magnitudes and perturbed them randomly using
Gaussian photometric errors inferred from the artificial star tests. We finally 
randomly chose whether a synthetic WD is observable using our determined completeness fraction as a function of the observed magnitudes.

The CMD of the synthetic objects that pass the completeness test (their number is similar to that of the observed sample) nicely overlaps the area covered by the observed WDs. 

%and shows an almost vertical morphology at its lower end, consistent --within the fairly large photometric errors-- with the observations, as it was the case with our previous analyses of NGC~6397 \citep{2024AN....34540039B} and 
%and M~4 \citep{2025arXiv250110070B} \textit{JWST} data.
%This morphology is different from the case of optical CMDs, where the presence of objects with increasing mass 
%(progeny of increasingly massive and shorter lived progenitors) 
%causes a turn to the blue at almost constant magnitude at 
%the bottom end of the isochrones (and observed CSs) 
%Due to the different behaviour of the bolometric corrections as a function of $T_{\rm eff}$ and surface gravity in these infrared filters, the turn to the blue in the optical is replaced by a continuous decrease of the magnitude towards slightly bluer colours. We will come back to this point later in the section.

\begin{centering} 
\begin{figure}
 \includegraphics[width=\columnwidth]{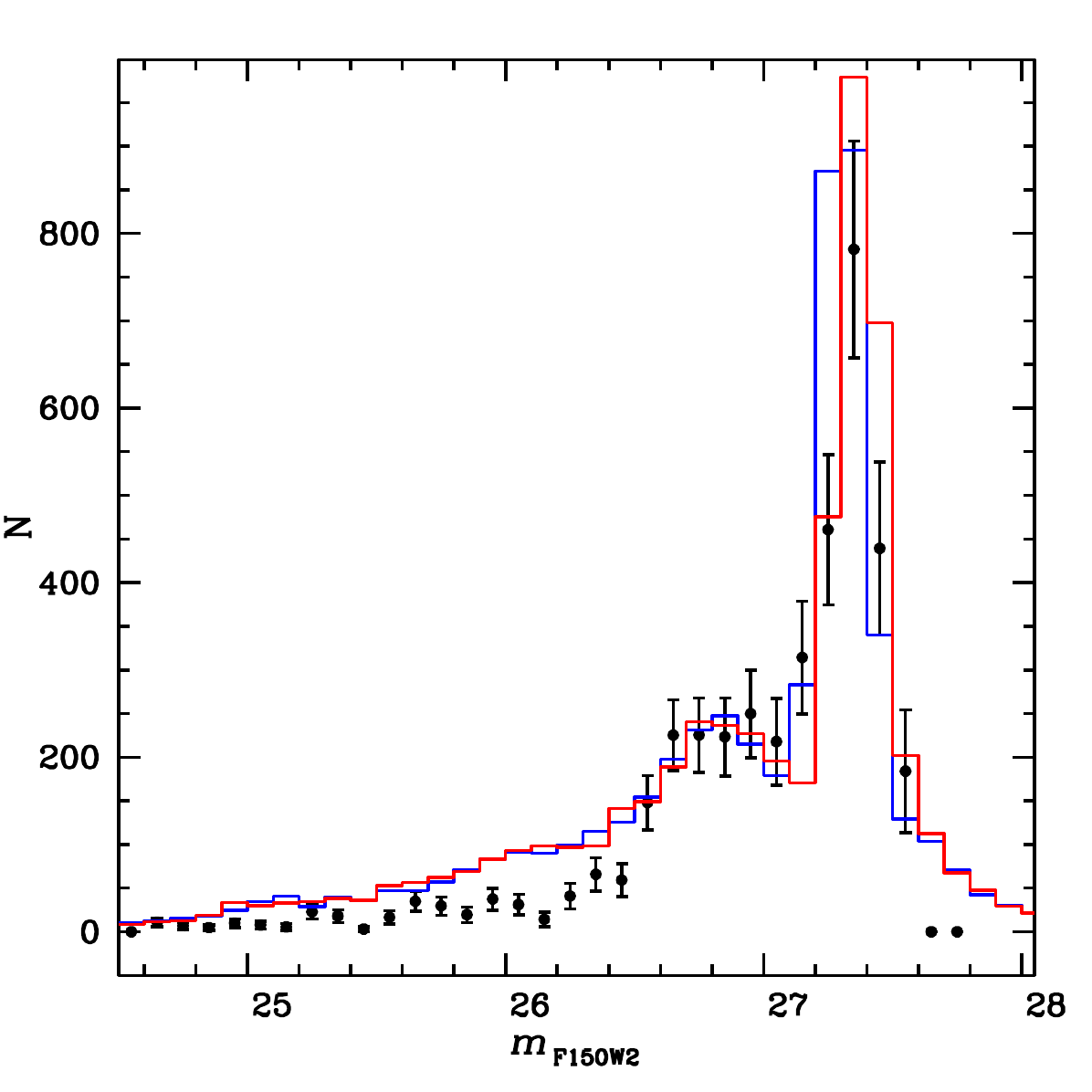}
 \caption{Comparison of the observed WD LF (filled circles with 
 error bars) and two theoretical LFs calculated for ages equal 
 to 11.5 (blue histogram) and 11.9~Gyr (red histogram -- see text for details).} 
 \label{lfage} 
\end{figure} 
\end{centering}

For a more quantitative determination of 47\,Tuc age from its cooling sequence  
we compared the completeness-corrected observed WD LF with its theoretical counterparts. 
The theoretical LFs have been obtained from the calculation of synthetic CMDs, as described above, to include the effect of the photometric errors. In the comparison  
with completeness-corrected LF we included the entire sample of synthetic objects, not just those that passed the completeness test.

Figure~\ref{lfage} displays the observed LF together with two theoretical 
LFs for ages equal to 11.5 and 11.9~Gyr, normalized to match the observed number 
of stars in the well-populated magnitude range $m_{\rm F150W2}$=26.6--26.9, where the star counts are still unaffected by the choice 
of the exact value of the cluster age. The theoretical LFs display a peak 
covering the three  
bins centred at $m_{\rm F150W2}$ between 27.25 and 27.45, 
and a sharp decrease at the same magnitudes as their observed counterpart. 
Ages outside this range produce LFs with the peak either brighter 
(age too young) or fainter (age too old) than the observed one.

\begin{centering} 
\begin{figure}
 \includegraphics[width=\columnwidth]{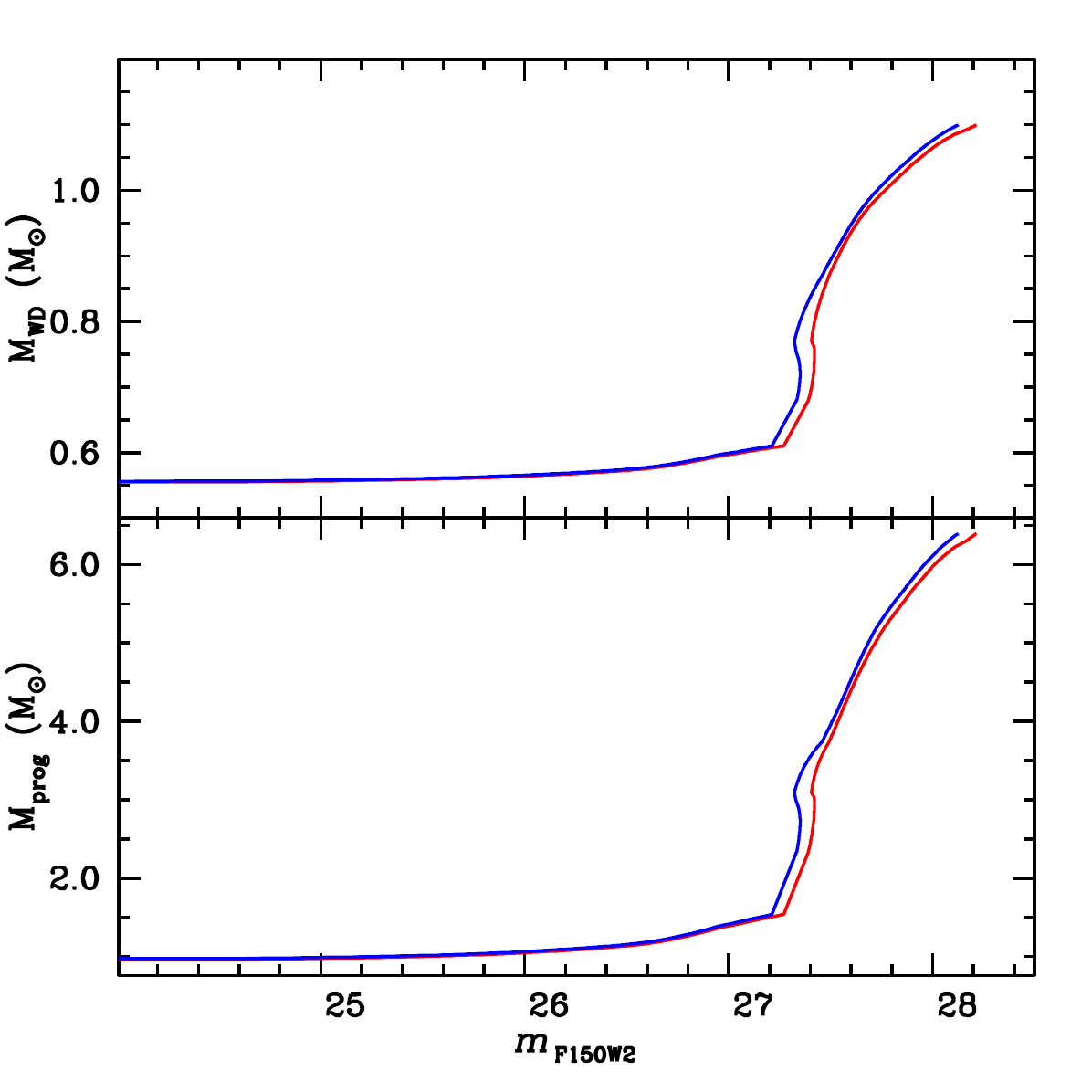}
 \caption{Values of the WD mass ($M_{\rm WD})$ and the 
 corresponding progenitor mass ($M_{\rm prog}$) as a function of $m_{\rm F150W2}$, along 
 the 11.5 (blue line) and 11.9~Gyr (red line) WD isochrones shifted to account for the cluster 
 extinction and distance modulus.} 
 \label{isomass} 
\end{figure} 
\end{centering}

We can also notice that the shape of 
the observed LF is not matched in its details by theory. In particular, at magnitudes brighter than the range used for the 
normalization, the predicted number counts are larger than 
the observations, especially in the 
range $m_{\rm F150W2} \sim$25.85--26.45. 
To investigate this issue further, we show in Fig.~\ref{isomass} 
the distribution of WD and progenitor masses along the 
isochrones employed to calculate the LFs in Fig.~\ref{lfage}.
Down to $m_{\rm F150W2}\sim$26.2--26.5, the 
isochrones  
are populated by WDs with essentially constant mass;  
indeed, in the range between $m_{\rm F150W2}\sim$25.0 and $m_{\rm F150W2}\sim$26.4 --corresponding to effective temperatures between 
$\sim$11,000~K and $\sim$5,500~K--the value of the WD mass $M_{\rm WD}$ ranges from 0.557-0.558$M_{\odot}$ 
to 0.570-0.572$M_{\odot}$, an almost negligible increase due to a 
combination of the short
cooling times (compared to the cluster age) at these luminosities and the almost constant 
value of the final WD mass for progenitors with initial 
masses up to $\sim 2M_{\odot}$ \citep[see][]{cummings}. 

It is important also to note that 
at $m_{\rm F150W2}$ fainter than $\sim$27.2~mag  
the mass of the WD (and progenitors) displays 
a very steep increase with increasing $m_{\rm F150W2}$. 
In the magnitude range of the observed narrow peak of the LF, the isochrones 
are populated by objects with masses between 
$\sim$0.6 and $\sim$0.95-1.0 ~$M_{\odot}$, encompassing almost the whole mass spectrum of CO WDs. This steep increase of the WD mass over a narrow magnitude increase near the faint end of the isochrone translates 
to a pile-up of objects in the LF, and 
produces the observed peak, which gets fainter with increasing age because 
of the increased cooling times of the WDs.

The property that the bright part of a globular cluster cooling sequence is 
populated by objects with virtually constant mass was exploited by 
\citet{empcool} to determine a semi-empirical cooling sequence 
for the bright WDs in 47\,Tuc, in the temperature range between $\sim$40,000~K and $\sim$7,000~K. After measuring the $T_{\rm eff}$ of a large sample of WDs, and assuming that the rate at which stars enter the cooling sequence is constant, \citet{empcool} 
employed stellar evolution models for the WD 
progenitors to determine the rate at which stars leave the main sequence, which must be the same as the rate at which objects enter the CS\footnote{The authors discounted the possibility 
that dynamical relaxation of the WDs could make this
assumption invalid, because the relaxation time evaluated for their field 
is too long compared to the time elapsed between the progenitors of their faintest observed WDs leaving the main sequence, and their WD progeny reaching their observed luminosity.}. 
The result of their analysis was a semi-empirical relation between 
$T_{\rm eff}$ and cooling times for the constant mass objects 
populating the bright WDs in 47\,Tuc.

\begin{centering} 
\begin{figure}
 \includegraphics[width=\columnwidth]{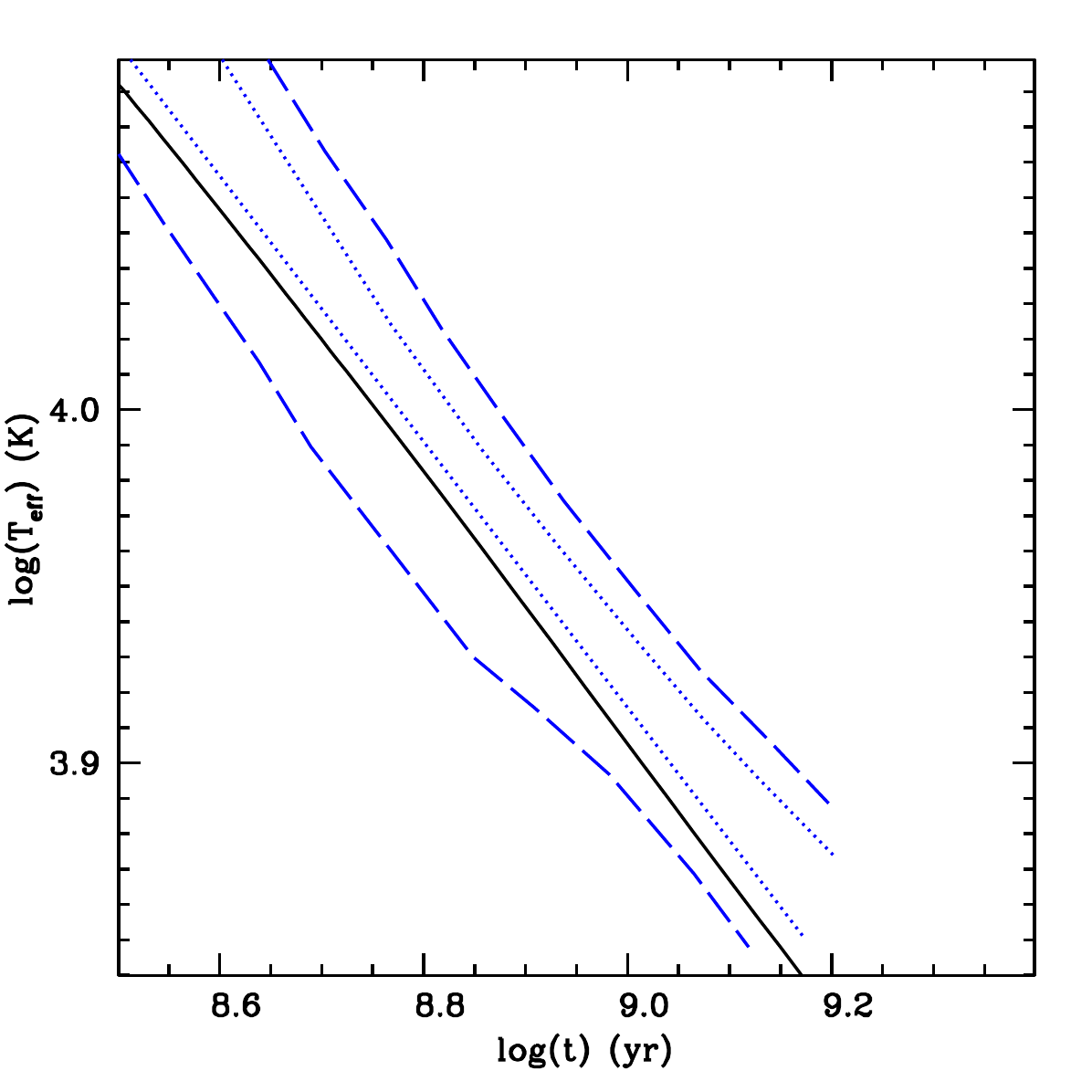}
 \caption{Semi-empirical cooling sequence ($T_{\rm eff}$ vs age) for 
 the bright WDs in 47\,Tuc \citep[from][]{empcool} compared to a theoretical 0.56$M_{\odot}$ counterpart (solid black line) from our adopted WD calculations. The dotted 
 blue lines enclose the 1$\sigma$ error range of the semi-empirical 
 sequence, whilst the dashed blue lines mark the 2$\sigma$ error 
 interval (see text for details).} 
 \label{empcool} 
\end{figure} 
\end{centering}

Figure~\ref{empcool} compares this semi-empirical cooling 
law with results from BaSTI-IAC WD calculations for a representative 0.56$M_{\odot}$ WD model. 
The highest temperature considered in the comparison 
corresponds to $m_{\rm F150W2}\sim$25 while the lowest temperature is at the cool 
limit of the semi-empirical calibration, that corresponds to 
$m_{\rm F150W2}\sim$26.
The figure shows that in this magnitude range the models are consistent 
with the semi-empirical calibration within less 
than 2$\sigma$, hence it is unlikely that the over-predicted star counts in the magnitude range between 25.85 
and 26.45 are due to an overestimate of the cooling times by the WD models.

% {\bf 
% The following paragraph (printed in italic) is just preliminary. Please check whether it makes sense. What is the relaxation time of this field? Can you calculate it? It is quite external.
% In principle, the more massive WDs at the end of the cooling sequence come from fast-evolving progenitors (up to a few hundred Myr lifetimes) whose mass distribution has not been affected by relaxation, but the WDs must have been affected if the relaxation time is comparable to their cooling age that is on the order of 10`Gyr or more. For the brighter constant-mass WDs instead the mass distribution of their long-lived progenitors might have been affected by dynamical relaxation.  
% Please check/comment/modify the paragraph if necessary. It is a very general statement}

% \textcolor{magenta}{R: 
% Possiamo anche sentire Vesperini se volete, 
% ma sia qui, che come nel caso di omega Cen, siamo ben oltre 1 $r\_h$, 
% dove il tempo di rilassamento e` gia` 10\,Gyr (Harris 1996, 2010 update). Fra 1.5 e quasi 3 mi aspetto che il tempo di rilassamento 
% sia molto maggiore dell'eta` dell'Universo. Sicche`, per quel che ne sappiamo forse 
% possiamo trascurare questi effetti. Le multi pop potrebbero avere altri effetti, ma mi hai sempre 
% spiegato che non dovrebbero esserci effetti sulla WDCS. Secondo me va bene 
% ed e` sufficiente quello che hai scritto qui sotto.}

The main reason for this mismatch, and 
the general inability of the theoretical LFs to match the exact star counts across the full magnitude range 
is likely because the WD mass 
distribution along the cooling sequence depends in a 
complex way on the progenitors' initial MF, and dynamical 
evolutionary effects (escape from the cluster of WD progenitors, mass segregation that has caused a spatial redistribution of the WDs and 
their progenitors) that have affected both the progenitors' and the WD mass distribution in the observed field \citep[see, e.g.,][]{richer}. 
We will go back to this point at the end of the section.

\begin{centering} 
\begin{figure}
 \includegraphics[width=\columnwidth]{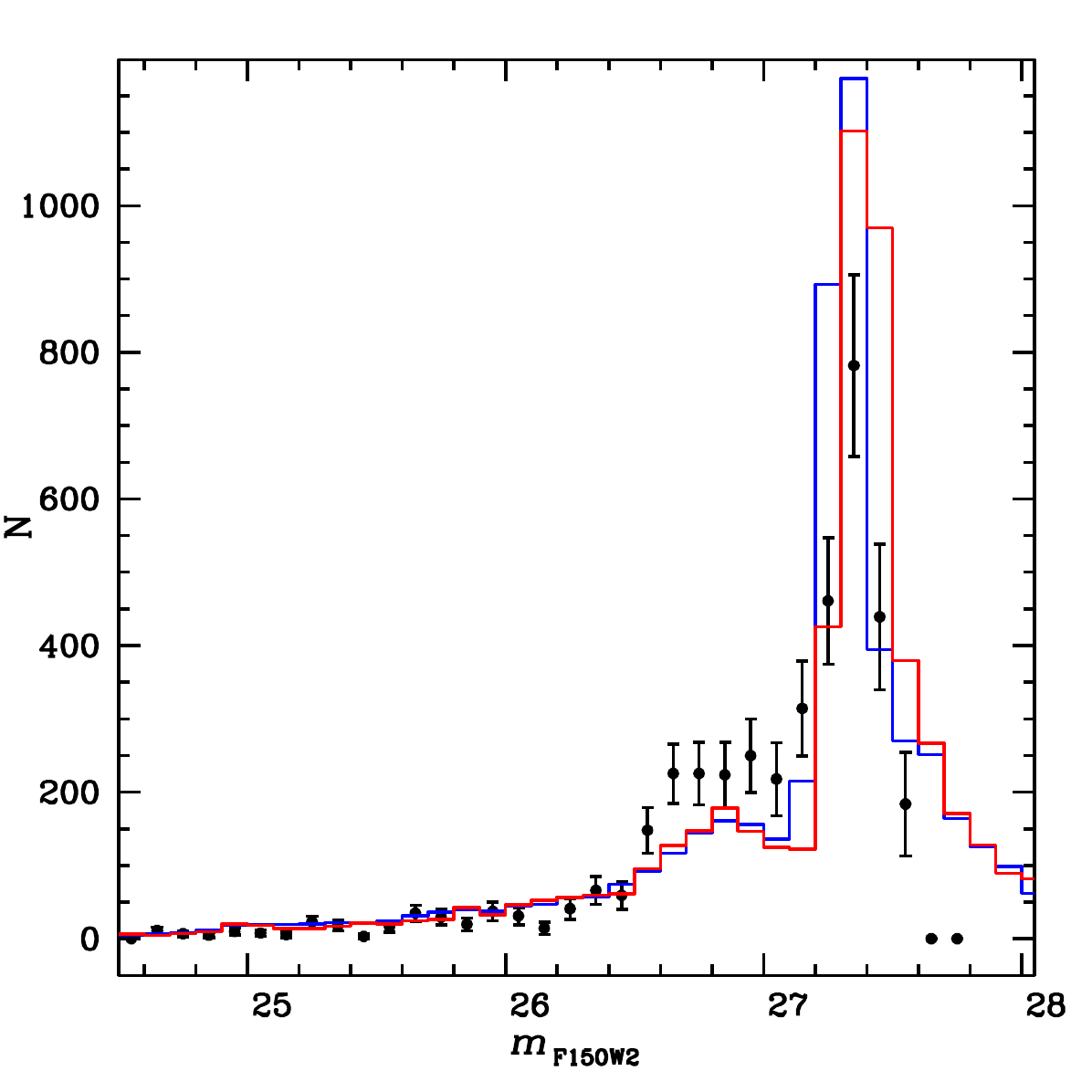}
 \caption{As Fig.~\ref{lfage} but the theoretical 
 LFs are calculated with a top-heavy MF for the 
 WD progenitors (see text for details).} 
 \label{lfageimf} 
\end{figure} 
\end{centering}

\begin{centering} 
\begin{figure}
 \includegraphics[width=\columnwidth]{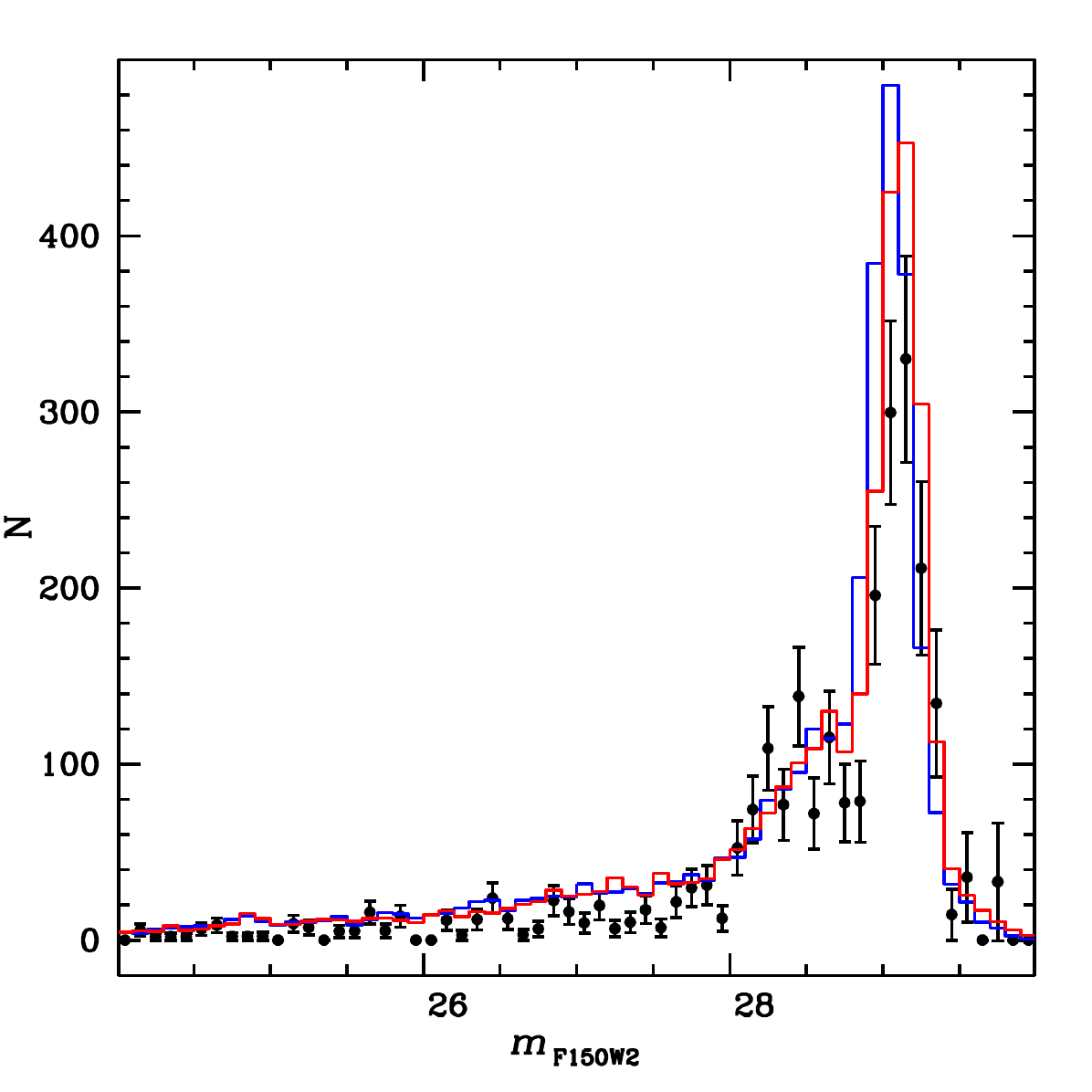}
 \caption{As Fig.~\ref{lfage} but the observed and theoretical 
 LFs are in the $HST$/ACS $F606W$ filter (see text for details).} 
 \label{lfageopt} 
\end{figure} 
\end{centering}

%\begin{centering} 
%\begin{figure}
% \includegraphics[width=\columnwidth]{figures/fig_paper_eclbin.eps}
% \caption{Mass-radius diagram for the components (filled circles with error bars) of the two eclipsing binary systems studied by \citet{eclbin}, and three pairs of colour-coded isochrones with the labelled [Fe/H] and ages. For each of the 
% selected metallicities, the two isochrones represent the 
% maximum age range compatible with the observations (see text for %details).} 
% \label{eclbin} 
%\end{figure} 
%\end{centering}

At this stage we show in Fig,~\ref{lfageimf} the effect of 
varying only the 
progenitors' MF exponent when comparing observed and theoretical LFs; 
changing this exponent does alter the mass WD mass distribution along the 
CS.
The 11.5 and 11.9~Gyr LFs shown in the figure  
have been calculated using a top-heavy progenitors' 
MF with an exponent equal to $-$1.3, without altering their normalization. 
This modification of the progenitor's MF increases the number of 
more massive WDs compared to the less massive ones, and the bright part of the observed LF is now reasonably well matched by theory, 
although there is an obvious mismatch at fainter magnitudes. 
A single exponent power law for the progenitors' MF clearly 
is not a good approximation however, for our purposes, the important point is that the magnitude range of the peak of the LF—the age indicator—is a solid prediction.
%To address this mismatch, a detailed analysis of the dynamical effects on the WD mass distribution in the observed field is needed (well beyond the goal of this work). However, for our purposes, the important point is that the magnitude range of the peak of the LF—the age indicator—is a solid prediction.

If we account for conservative errors of $\pm$0.05~mag on the 
eclipsing binary distance \citep[considering the spread of the individual components' distances determined with the two methods discussed in][]{eclbin}, $\pm$0.1~dex on [Fe/H], and $E(B-V)$ ranging from the reference $E(B-V)$=0.024 value taken 
from \citet{gratton} to the more standard $E(B-V)$=0.04 \citep[see, e.g.,][]{koch}, the total age range compatible with the position of 
the peak of the observed LF ranges between 11.3 to 12.3~Gyr. The contributions of the uncertainties on the reddening and 
[Fe/H] is negligible compared to the effect of the uncertainty on the distance modulus.

%\subsection{Comparison with the optical LF and the age from the cluster eclipsing binaries}

%{\bf subsection to be eliminated. Keep only comparison with optical LF %and move it in the main section}

As a further step in our analysis, we have checked the consistency between the cluster age 
determined from \textit{JWST} and \textit{HST} observations. 
Figure~\ref{lfageopt} compares the completeness-corrected optical LF with theoretical LFs for 11.5 and 11.9~Gyr (as in Fig.~\ref{lfage}), calculated with the same progenitors' MF, the reference distance modulus and reddening employed in Fig.~\ref{lfage}, extinction  
$A_{\rm F606W}$ computed following \citet{2005MNRAS.357.1038B}, 
and photometric errors obtained from the artificial star analysis 
of the $HST$ observations.
The theoretical LFs have been rescaled to match the observed number of stars between $m_{\rm F606W}$=28.05 and 28.65, corresponding approximately to the same magnitude range of the \textit{JWST} LF normalization. As we found and discussed for the infrared data, 
there is 
on average an excess of objects at magnitudes brighter than the 
normalization interval, and more in general the theoretical LFs 
do not match the detailed shape of the observed counterpart. 
The position of the peak in the observed LF sensitive to age is 
nicely bounded by the two theoretical LFs, as in the case 
of the \textit{JWST} data, confirming the consistency between the ages obtained 
from infrared and optical observations of the CS.

%Finally, Fig.~\ref{eclbin} shows the total age range compatible 
%with the masses and radii measured for the two cluster eclipsing binaries studied by \citet{eclbin}. We compare here the data with theoretical $\alpha$-enhanced isochrones \citep[from][]{bastiiacaen}, with the underlying assumption that the four objects have to be coeval, with an age equal to the age of the cluster.
%Taking into account the uncertainty of the cluster [Fe/H] ($\pm$0.1~dex) and the error bars on the empirical masses and radii, the figure confirms that  
%the age range compatible with the \textit{JWST} CS LF (11.3 to 12.3~Gyr) 
%if fully consistent with the eclipsing binary data.

\subsection{The impact of the cluster multiple populations}

It is well established that 47\,Tuc stars display a range of initial helium abundances. 
Indeed, analyses of the cluster CMD 
(main sequence, subgiant branch, RGB, horizontal branch) by several groups 
\citep[see, e.g.,][]{apkb,dicriscienzo,hbsim,milon} have disclosed a range of initial He mass fractions $\Delta Y$=0.02-0.03. This spread of initial He abundance is a 
manifestation of the presence of  
multiple, roughly coeval stellar populations in the cluster, a common 
occurrence in massive star clusters like 
the Galactic GCs \citep[see, e.g.][and references therein]{gcb,bl18, cs20, mm}.

In our adopted progenitor models a 
metallicity [Fe/H]=$-$0.7 corresponds to a standard $Y$=0.255 \citep[see][for details]{bastiiacaen}, 
hence the initial helium spread in the cluster can be taken into account 
by considering models with $Y$ between 0.255 and 0.275-0.285.
We have verified 
using models in the BaSTI-IAC database \citet{bastiiacaen}
and by computing appropriate test calculations with the BaSTI-IAC code, 
that for $\Delta Y$=0.03 the variation of the initial mass of TO stars (at fixed age) in [Fe/H]=$-$0.7 $\alpha$-enhanced old isochrone with ages 
between 10 and 12~Gyr equals just $\sim$0.03~$M_{\odot}$ (the mass is reduced when the initial helium increases).
In the assumption that the WD initial-final mass relation 
\citep[we remind the reader that we employ the semi-empirical relation derived by][]{cummings} is not affected by this small variation of initial helium, the mass of the objects evolving 
along the bright part of the corresponding WD isochrone is 
essentially unchanged compared to the standard $Y$=0.255 case. 
As for the more massive WDs in the region of the peak of the LF, the variation of their progenitor lifetime due to a small non-zero $\Delta Y$ 
is negligible compared to 
the cooling age at those magnitudes. As a result, the presence of 
a small $Y$ spread among 47\,Tuc stars should not have affected appreciably 
our age determination described in this section. 

The presence of multiple populations in the cluster, however, might have an impact on the detailed shape of the WD LF. 
Multiple stellar populations, potentially characterized by different initial stellar MFs \citep[e.g.,][]{2024AN....34540018S}, tend to exhibit distinct spatial distributions, as observed in other massive globular clusters like $\omega$\,Centauri \citep[e.g.,][]{2009A&A...507.1393B,2024A&A...688A.180S}. These spatial variations reflect the different kinematical properties of the medium from which the various stellar populations formed \citep[e.g.,][]{2015ApJ...810L..13B}, and they 
are an additional reason why a simple power-law description of the 
WD progenitor MF used in the calculation of the theoretical WD LFs is very 
likely not appropriate to match the exact shape of the observed LFs.

\section{Summary and conclusions}\label{conclusions}

%{\bf Please add a brief summary about the data}
%\textcolor{magenta}{R: ok, provvediamo ASAP!\\  
%L'abstract chi lo scrive?}
%
%\textbf{
%
%
%\textcolor{magenta}{\textbf{R: Michele, please improve this paragraph, complete %it with 
%all the observational results.}}
In this work, we studied the WD cooling sequence of the GC 47\,Tuc. We employed recent deep infrared observations from \textit{JWST} \citep[GO-2559;][]{2021jwst.prop.2559C}, obtained in $\sim$2022.7 with NIRCam’s ultra-wide filters \citep[see][for a complete description of the dataset]{2025A&A...694A..68S}, and ultra-deep optical imaging from \textit{HST} \citep[GO-11677;][]{2009hst..prop11677R} of an overlapping field, taken with ACS in $\sim$2010.4.

Thanks to the $\sim$12-year time baseline, we measured precise PMs, 
which allowed us to isolate a clean sample of cluster members. This selection effectively separated cluster WDs from foreground Galactic field stars, background SMC members, and unresolved distant galaxies.

Using this cleaned WD sample, we derived the LF of the WD cooling sequence in both \textit{JWST} infrared and \textit{HST} optical filters. The two LFs exhibit a similar shape, peaking at $m_{\rm F150W2} \sim 27.5$ ($m_{\rm F606W} \sim 29.1$).

We have then determined the age of the cluster by employing the 
\textit{JWST} cooling sequence LF, and obtained a value of 11.8$\pm$0.5~Gyr, where the error 
includes the uncertainty due to matching the magnitude range of the 
peak of the LF and the uncertainty on the eclipsing binary distance modulus \citep{eclbin} assumed for the cluster (the uncertainties on the cluster metallicity and reddening give a negligible contribution).
In the process, we have tested the cooling times of the WD model (with a mass of 0.56$M_{\odot}$) populating the bright part of the cluster cooling sequence by comparison with the semi-empirical calibration by \citet{empcool}, based on observations of the bright WDs 47\,Tuc. We found that the model agrees with this calibration within less than 2$\sigma$.

We have also found that the age derived from the 
infrared cooling sequence LF is  
consistent with the optical LF obtained from the \textit{HST} observations. 
%Very importantly, 
%it is also consistent with the age derived from matching the position of the two cluster's eclipsing binaries studied by \citet{eclbin} with theoretical isochrones (consistent with the models used to calculate the WD progenitors' lifetimes) in the mass-radius diagram. 

When comparing with the previous 47\,Tuc cooling sequence analyses by \citet{47tuchansen} and \citet{47tucgb} that made use of 
optical \textit{HST} data, our age estimate is higher than the 
9.9$\pm$0.7~Gyr (95\% confidence level) determined by \citet{47tuchansen} using different cooling models \citep{hansen99}, despite a distance modulus that is almost the same (larger by just 0.03~mag) as the one employed in our analysis. 
Our result is instead consistent with \citet{47tucgb} who 
employed the \citet{renedo} WD models and 
determined 12.5$\pm$0.5 and 12.5$\pm$1.0, depending on the method 
used (there is no information in the paper about the distance modulus).

Regarding the study by \citet{campos}, who made use of \citet{romero} WD models, our age estimate is consistent with their age obtained using the distance ($(m-M)_0$=13.26, just 0.03~mag shorter than our adopted eclipsing binary distance) and reddening ($E(B-V)$=0.023) derived from fitting models to the main sequence \citep{dotterdist}.  
When fitting the bright part of the cooling sequence to models, the higher derived reddening ($E(B-V)$=0.045) and distance modulus (0.02`mag longer) combine  to decrease the age estimate and explain the difference 
of the two different ages estimated by \citet{campos}.

Finally, our cluster age determined from the cooling sequence in the infrared confirms 47\,Tuc age derived from the cluster main sequence TO \citep[11.75$\pm$0.25~Gyr as determined by][]{vdbages} and by the joint analysis of the TO and the mass-radius diagram of two eclipsing binaries in the cluster \citep[12.0$\pm$0.5~Gyr as determined by][]{eclbin}, 
and the location of 47\,Tuc within the body of the age-metallicity relation for in situ GCs \citep[see, e.g.,][]{leaman}. 

\section*{Acknowledgements}

We dedicate this paper to the memory of our colleague Prof. Harvey Richer ($\star$ April 1944 -- $\dagger$ 13 November 2023), a highly accomplished astronomer expert in stellar populations and in particular globular clusters, who passed away during the development of this project. His focus was the late stages of stellar evolution, particularly carbon stars and white dwarfs. We thank Enrico Vesperini for discussions about the dynamical evolution of globular clusters.
We thank our Referee, Kurtis Williams, for a constructive 
report that has improved the presentation of our results.
SC acknowledges financial support from PRIN-MIUR-22: CHRONOS: adjusting the clock(s) to unveil
the CHRONO-chemo-dynamical Structure of the Galaxy” (PI: S. Cassisi) granted by the European Union - Next Generation EU,
and the support of a fellowship from La Caixa Foundation (ID 100010434) with fellowship code LCF/BQ/PI23/11970031
(P.I.: A. Escorza) and from the Fundación Occident and the Instituto de Astrofísica de Canarias under the
Visiting Researcher Programme 2022-2025 agreed between both institutions.
%%%

\section*{Data Availability}
The WD models employed in this work are available at \url{http://basti-iac.oa-abruzzo.inaf.it/}. The LFs are  
available at
\url{https://web.oapd.inaf.it/bedin/files/PAPERs_eMATERIALs/JWST/Paper_47Tuc_WDCS/}.

\bibliographystyle{mnras}
\bibliography{main}

\appendix

\section{Colour-magnitude diagrams of the white dwarf cooling sequence of 47\,Tuc}

\begin{centering} 
\begin{figure*}
 \includegraphics[width=\textwidth]{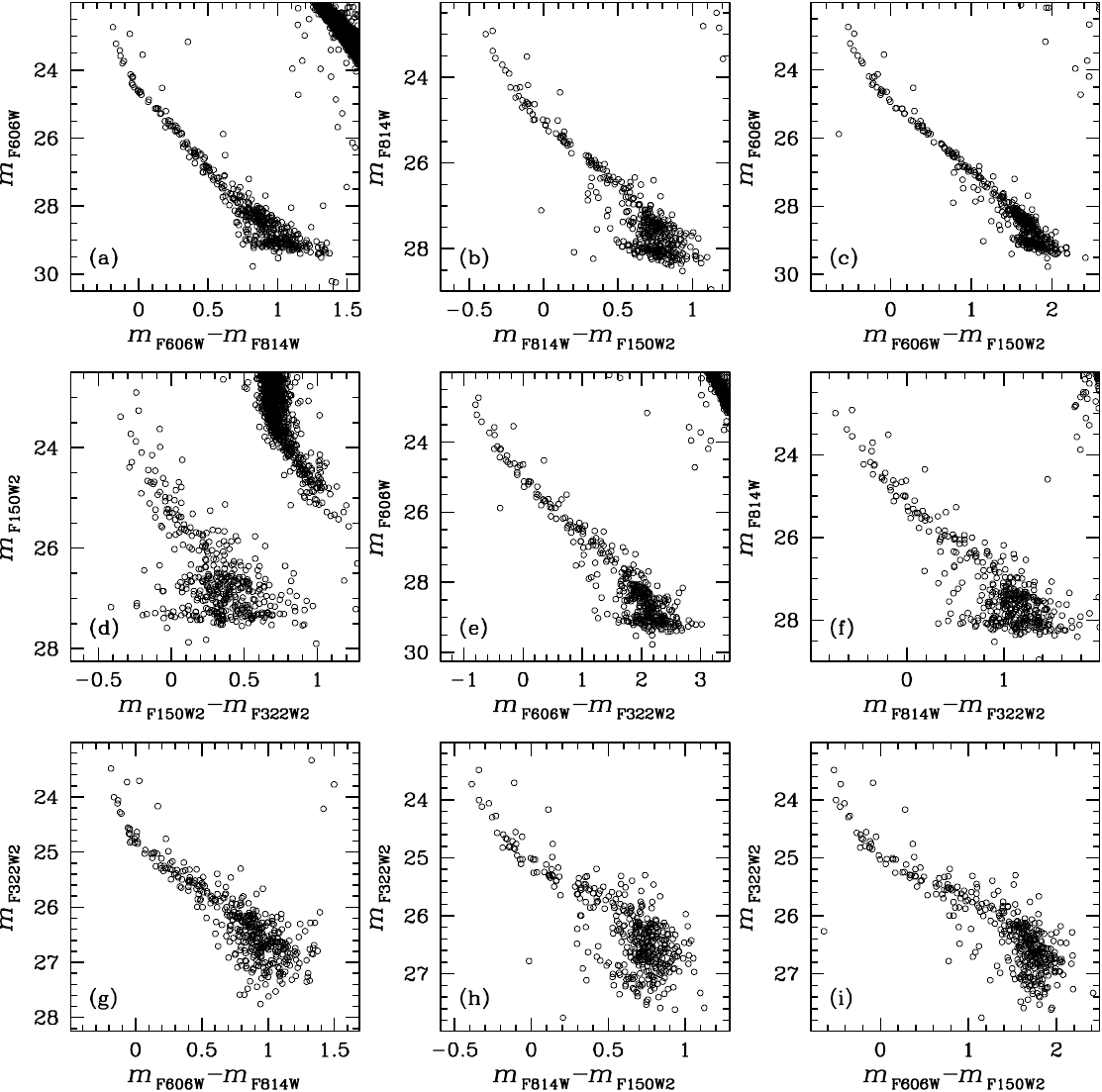}
 \caption{Nine CMDs of 47\,Tuc  WD CS, using the four \textit{HST} and \textit{JWST} filters employed in this study.} 
 \label{cmd} 
\end{figure*} 
\end{centering}

We show in Fig.~\ref{cmd} a set of CMDs of the WD cooling sequence of 47\,Tuc, using various combinations of the \textit{HST} and \textit{JWST} filters employed in this study. The CMDs include the well-measured sources that passed the PM selection described in Fig.\,\ref{pm}.

\section{Luminosity functions of the cooling sequence of 47\,Tuc }

We present in Tables~\ref{tab:jwst} and ~\ref{tab:hst} 
the LFs of the WD cooling sequence from \textit{JWST} and \textit{HST} data, used for the 
cluster age determination.

\begin{table}
    \centering
        \caption{Completeness-corrected infrared luminosity function of 47\,Tuc WDs derived from \textit{JWST} photometry.}
    \begin{tabular}{cccccc}
        \cmidrule(lr){1-3} \cmidrule(lr){4-6}
        $m_{\rm F150W2}$ & ${\rm N}_{\rm stars}$ & $\delta{\rm N}_{\rm stars}$ & $m_{\rm F150W2}$ & ${\rm N}_{\rm stars}$ & $\delta{\rm N}_{\rm stars}$ \\
        \cmidrule(lr){1-3} \cmidrule(lr){4-6}
        22.05 & 0.00  & 0.00  & 24.95 & 9.93   & 4.97   \\
        22.15 & 0.00  & 0.00  & 25.05 & 7.78   & 4.50   \\
        22.25 & 0.00  & 0.00  & 25.15 & 5.43   & 3.84   \\
        22.35 & 0.00  & 0.00  & 25.25 & 22.81  & 8.08   \\
        22.45 & 0.00  & 0.00  & 25.35 & 17.92  & 7.33   \\
        22.55 & 0.00  & 0.00  & 25.45 & 3.14   & 3.14   \\
        22.65 & 0.00  & 0.00  & 25.55 & 16.51  & 7.40   \\
        22.75 & 0.00  & 0.00  & 25.65 & 34.85  & 11.06  \\
        22.85 & 0.00  & 0.00  & 25.75 & 29.52  & 10.47  \\
        22.95 & 1.56  & 1.56  & 25.85 & 19.54  & 8.75   \\
        23.05 & 0.00  & 0.00  & 25.95 & 37.36  & 12.50  \\
        23.15 & 0.00  & 0.00  & 26.05 & 31.00  & 11.75  \\
        23.25 & 1.65  & 1.65  & 26.15 & 14.23  & 8.23   \\
        23.35 & 1.67  & 1.67  & 26.25 & 40.88  & 14.51  \\
        23.45 & 0.00  & 0.00  & 26.35 & 65.83  & 19.13  \\
        23.55 & 0.00  & 0.00  & 26.45 & 59.21  & 18.83  \\
        23.65 & 1.73  & 1.73  & 26.55 & 147.91 & 31.29  \\
        23.75 & 1.75  & 1.75  & 26.65 & 225.19 & 40.67  \\
        23.85 & 1.78  & 1.78  & 26.75 & 225.33 & 42.75  \\
        23.95 & 1.82  & 1.82  & 26.85 & 223.33 & 44.72  \\
        24.05 & 1.86  & 1.86  & 26.95 & 249.66 & 50.09  \\
	24.15 & 1.90  & 1.90  & 27.05 & 217.72 & 49.64  \\
	24.25 & 3.88  & 2.74  & 27.15 & 314.15 & 64.60  \\
	24.35 & 2.00  & 2.00  & 27.25 & 460.62 & 86.13  \\
	24.45 & 4.13  & 2.92  & 27.35 & 781.79 & 124.19 \\
	24.55 & 0.00  & 0.00  & 27.45 & 439.15 & 99.27  \\
	24.65 & 11.03 & 4.94  & 27.55 & 183.82 & 70.50  \\
	24.75 & 6.86  & 3.96  & 27.65 & 0.00   & 0.00   \\
        24.85 & 4.76  & 3.37  & 27.75 & 0.00   & 0.00   \\
        \cmidrule(lr){1-3} \cmidrule(lr){4-6}
    \end{tabular}
    \label{tab:jwst}
\end{table}

\begin{table}
    \centering
        \caption{Completeness-corrected optical luminosity function of 47\,Tuc WDs derived from \textit{HST} photometry.}
    \begin{tabular}{cccccc}
        \cmidrule(lr){1-3} \cmidrule(lr){4-6}
        $m_{\rm F606W}$ & ${\rm N}_{\rm stars}$ & $\delta{\rm N}_{\rm stars}$ & $m_{\rm F606W}$ & ${\rm N}_{\rm stars}$ & $\delta{\rm N}_{\rm stars}$ \\
        \cmidrule(lr){1-3} \cmidrule(lr){4-6}
        22.65  & 0.00  & 0.00 & 26.35 &  11.72 &  5.88 \\
        22.75  & 1.59  & 1.59 & 26.45 &  23.93 &  8.52 \\
        22.85  & 0.00  & 0.00 & 26.55 &  12.22 &  6.13 \\
        22.95  & 1.61  & 1.61 & 26.65 &   3.12 &  3.13 \\
        23.05  & 0.00  & 0.00 & 26.75 &   6.38 &  4.52 \\
        23.15  & 0.00  & 0.00 & 26.85 &  22.49 &  8.56 \\
        23.25  & 1.64  & 1.64 & 26.95 &  16.17 &  7.27 \\
        23.35  & 0.00  & 0.00 & 27.05 &   9.76 &  5.65 \\
        23.45  & 1.69  & 1.69 & 27.15 &  19.65 &  8.07 \\
        23.55  & 3.44  & 2.44 & 27.25 &   6.59 &  4.67 \\
        23.65  & 0.00  & 0.00 & 27.35 &  10.12 &  5.86 \\
        23.75  & 3.56  & 2.52 & 27.45 &  17.26 &  7.76 \\
        23.85  & 0.00  & 0.00 & 27.55 &   7.07 &  5.01 \\
        23.95  & 0.00  & 0.00 & 27.65 &  21.73 &  8.93 \\
        24.05  & 0.00  & 0.00 & 27.75 &  29.71 & 10.59 \\
        24.15  & 5.81  & 3.36 & 27.85 &  31.27 & 11.15 \\
        24.25  & 1.98  & 1.98 & 27.95 &  12.38 &  7.17 \\
        24.35  & 2.01  & 2.01 & 28.05 &  52.42 & 15.32 \\
        24.45  & 2.05  & 2.05 & 28.15 &  74.26 & 18.88 \\
        24.55  & 6.25  & 3.62 & 28.25 & 108.91 & 23.77 \\
        24.65  & 8.49  & 4.25 & 28.35 &  76.95 & 20.19 \\
        24.75  & 2.16  & 2.16 & 28.45 & 138.39 & 27.92 \\
        24.85  & 2.21  & 2.21 & 28.55 &  71.89 & 20.23 \\
        24.95  & 2.25  & 2.25 & 28.65 & 115.10 & 26.33 \\
        25.05  & 0.00  & 0.00 & 28.75 &  77.98 & 21.96 \\
        25.15  & 9.40  & 4.71 & 28.85 &  78.79 & 23.10 \\
        25.25  & 7.21  & 4.17 & 28.95 & 195.79 & 39.13 \\
        25.35  & 0.00  & 0.00 & 29.05 & 299.60 & 52.16 \\
        25.45  & 5.02  & 3.56 & 29.15 & 329.97 & 58.65 \\
        25.55  & 5.14  & 3.64 & 29.25 & 211.19 & 49.33 \\
        25.65  & 15.77 & 6.47 & 29.35 & 134.47 & 41.70 \\
        25.75  & 5.39  & 3.81 & 29.45 &  14.51 & 14.56 \\
        25.85  & 13.63 & 6.12 & 29.55 &  35.71 & 25.45 \\
        25.95  & 0.00  & 0.00 & 29.65 &   0.00 &  0.00 \\
        26.05  & 0.00  & 0.00 & 29.75 &  33.11 & 33.40 \\
        26.15  & 11.33 & 5.68 & 29.85 &   0.00 &  0.00 \\
        26.25  & 2.87  & 2.87 & 29.95 &   0.00 &  0.00 \\
        \cmidrule(lr){1-3} \cmidrule(lr){4-6}
    \end{tabular}
    \label{tab:hst}
\end{table}

\section{He-atmosphere WDs}

Prompted by our Referee, we have investigated whether the presence of helium-atmosphere WDs could change the age estimated  
by considering a pure hydrogen-atmosphere population.

\begin{centering} 
\begin{figure}
 \includegraphics[width=\columnwidth]{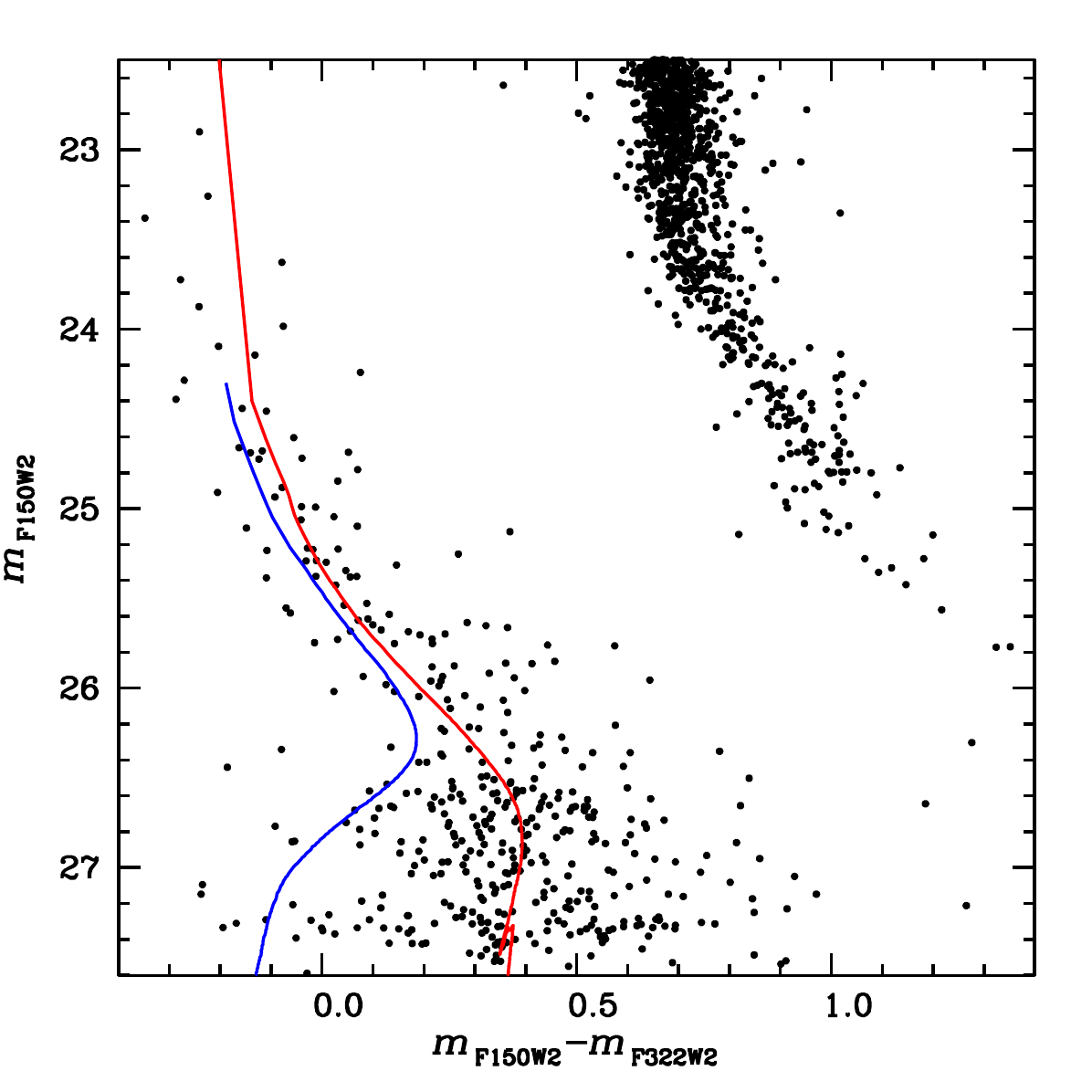}
 \caption{CMD of the 47~Tuc cooling sequence and very-low main sequence, together 
with a 12 Gyr hydrogen-atmosphere WD isochrone (red line --the same used to produce the 
synthetic WD population shown in Fig.~\ref{cmdiso}) and 
a 12~Gyr helium-atmosphere WD isochrone (blue line). The isochrones 
are shifted in magnitude and colour to account for the distance and reddening of the cluster.} 
 \label{cmdisoDB}
\end{figure} 
\end{centering}

Figure~\ref{cmdisoDB} compares the $JWST$ CMD of the 
cluster cooling sequence with the 12~Gyr hydrogen-atmosphere isochrone used to produce the 
synthetic WD population shown in Fig.~\ref{cmdiso}, and 
the helium-atmosphere counterpart calculated using 
the helium-atmosphere BaSTI-IAC WD calculations \citep{bastiiacwd}.
Both isochrones are shifted in magnitude and colour to account for 
the cluster distance and reddening.

Due to the faster cooling times, the termination of 
the helium-atmosphere isochrone is located 
at magnitudes much fainter than the faint limit of our photometry.
In the magnitude range of the observed cooling sequence, the 
mass of the objects populating the He-atmosphere 
isochrone spans a very narrow range, between $\sim$0.54 and $\sim$0.59 $M_{\odot}$.

This infrared diagram well separates the two classes of objects at $m_{\rm F150W2}$ fainter than $\sim$26.4 where, due to the effect of the different chemical composition on the 
bolometric corrections to these filters, the 
helium-atmosphere isochrone lies approximately at the blue end of 
the observed sequence.

Taken at face value, this comparison seems to confirm that helium-atmosphere objects 
should make only a small percentage (if any) of the observed 
cluster WD population.

\begin{centering} 
\begin{figure}
 \includegraphics[width=\columnwidth]{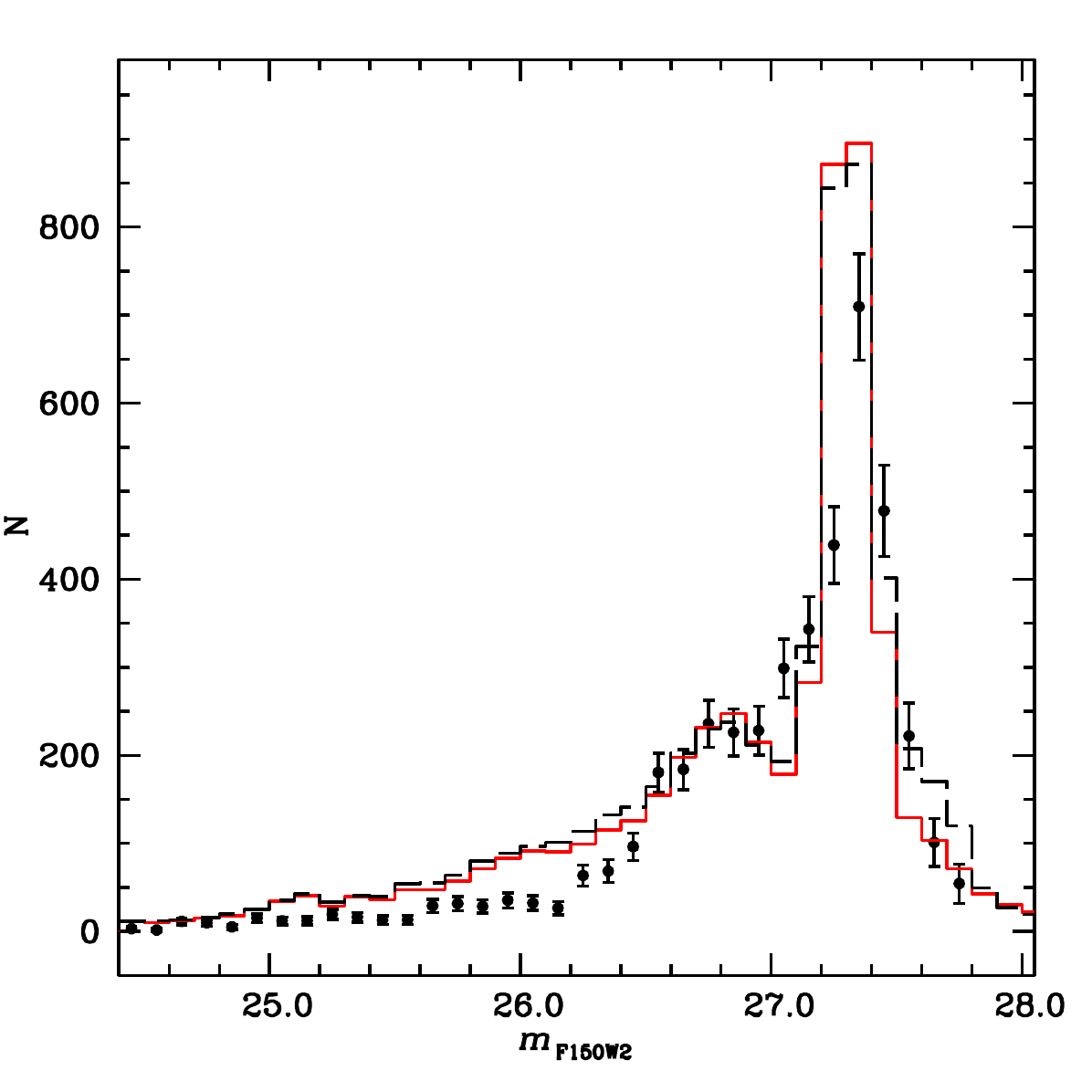}
 \caption{Two 11.5~Gyr WD LFs calculated as described in Sect.~\ref{age}, compared to the observed one. 
 The solid red line displays the LF 
 shown in Fig.~\ref{lfage}, that considers only hydrogen-atmosphere WD models; the dashed line displays a LF calculated for 
 a population with 75\% hydrogen-atmosphere, and 25\% helium-atmosphere objects, respectively. Both LFs are normalized as described in Sect.~\ref{age}.} 
 \label{lfDADB} 
\end{figure} 
\end{centering}

Nevertheless, we tested the effect of a sizable 25\% fraction of 
helium-atmosphere objects along the observed CS 
on the theoretical LFs used for our age 
determination. The result is shown in Fig.~\ref{lfDADB}, 
which compares the 11.5~Gyr LF that includes only hydrogen-atmosphere WDs, and a counterpart calculated for 
 a population with 75\% hydrogen-atmosphere and 25\% helium atmosphere objects in the magnitude interval covered 
 by the hydrogen-atmosphere isochrone. Both LFs are normalized as described in Sect.~\ref{age}.

 The inclusion of the helium-atmosphere objects barely changes the shape of the LF, and the effect on the age 
 determination from the cooling sequence is negligible.

\bsp
\label{lastpage}
\end{document}